\documentclass[aps,prb,twocolumn,showpacs,amsmath,amssymb,superscriptaddress,10pt]{revtex4-1}

\usepackage[utf8]{inputenc}

\usepackage{graphicx}
\usepackage[colorlinks=true,citecolor=blue]{hyperref}
\usepackage{units}
\usepackage{color}

\usepackage{verbatim} 

\renewcommand{\vec}[1]{\mathbf{#1}}

\DeclareMathOperator{\T}{T}

\def\up{\mathord{\uparrow}}
\def\down{\mathord{\downarrow}}

\begin{document}

\title{Odd-triplet superconductivity in single-level quantum dots}

\author{Stephan Weiss}
\affiliation{Theoretische Physik, Universit\"at Duisburg-Essen and CENIDE, 47048 Duisburg, Germany}
\author{J\"urgen K\"onig}
\affiliation{Theoretische Physik, Universit\"at Duisburg-Essen and CENIDE, 47048 Duisburg, Germany}

\date{\today}

\begin{abstract}
We study the interplay of spin and charge coherence in a single-level quantum dot.
A tunnel coupling to a superconducting lead induces superconducting correlations in the dot.
With full spin symmetry retained, only even-singlet superconducting correlations are generated.
An applied magnetic field or attached ferromagnetic leads partially or fully reduce the spin symmetry, and odd-triplet superconducting correlations are generated as well. 
For single-level quantum dots, no other superconducting correlations are possible.
We analyze, with the help of a diagrammatic real-time technique, the interplay of spin symmetry and superconductivity and its signatures in electronic transport, in particular current and shot noise.
\end{abstract}


\maketitle

\section{Introduction}

The formation of singlet Cooper pairs in conventional superconductors is a consequence of the broken $U(1)$-gauge symmetry in the electronic system below the critical temperature\cite{bcs}.  
The $SU(2)$-spin symmetry, on the other hand, remains intact in a BCS (Bardeen-Cooper-Schrieffer) superconductor.
The combination of superconductivity and magnetism, however, allows for other, unconventional, types of pairing.
Bulk rare-earth metals can behave as ferromagnetic superconductors\cite{Fertig, Ishikawa}.
Heavy-fermion and organic superconductors in strong magnetic fields can accommodate FFLO (Fulde-Ferrell-Larkin-Ovchinnikov) states\cite{Fulde, Larkin}, superconducting states with oscillating order parameter.
More recently, heterostructures built from conventional superconducting and ferromagnetic metals have become an active research field.
Theoretical investigations involving wavefunction-\cite{Halterman,Fritsch,Eschrig_rep} or semiclassical \cite{leggett_theoretical_1975, bergeret_rmp,Eschrig_rep} methods have predicted that the proximity of conventional superconductors leads to triplet pairing in the ferromagnets.
This has been experimentally confirmed in a variety of different heterostructures\cite{Ryazanov,Kontos2002}, in part by observing $0$-$\pi$-transitions in Josephson junctions\cite{robinson,keizer,birge}.
To establish a long-range superconducting coherence in the ferromagnet, it is necessary to break any residual uniaxial spin symmetry, e.g., by magnetic inhomogeneities\cite{bergeret_rmp} or spin-orbit coupling\cite{Bergeret_conv}. 

Superconducting correlations can be classified by their symmetry in spin, space, and time. 
Since electrons are fermions, the correlation function has to be antisymmetric under exchange of two particles.
This allows for four different classes of correlations.
First, we distinguish between correlators that are even or odd in time (or, equivalently, in frequency).
Second, each of them can be characterized by their spin symmetry, which may be even for triplets or odd for singlets.
Then, the symmetry in space is fixed. 
We denote the four classes of correlations as (i) even singlet, (ii) even triplet, (iii) odd triplet, and (iv) odd singlet, where "even" and "odd" refer to the symmetry in time (frequency) and "singlet" and "triplet" to spin.
(i) Even-singlet correlations appear in conventional $s$-wave superconductors but also in $d$-wave superconductors~\cite{tsuei_pairing_2000}. 
(ii) Even-triplet pairing is possible for $p$-wave superconductors, as suggested for Sr$_2$RuO$_4$~\cite{mackenzie_superconductivity_2003}.
(iii) Odd-triplet Cooper pairs have been demonstrated to deeply penetrate into ferromagnets.
In such a way, it was possible to control Josephson junctions\cite{Klose,Martinez} including $0$-$\pi$-transitions\cite{Gingrich_2015} via the magnetic configuration of the heterostructure, which gave rise to the term {\it superconducting spintronics}\cite{super_spintronics}.
Thereby, not only ferromagnetic metals but also ferromagnetic insulators have been used\cite{senapati}. Based on theoretical investigations it has been shown that odd-frequency triplet pairing also appears in diffusive normal metals contacted by an even-frequency triplet superconductor\cite{Tanaka}.
(iv) Odd-singlet superconductivity has only been predicted theoretically\cite{Balatsky} but not found experimentally yet.

In bulk systems, the possibility to influence the pairing mechanism by tuning external parameters is rather limited. 
For the heterostructures studied in Refs.~\onlinecite{Klose,Gingrich_2015,Martinez}, external magnetic fields have been used to change the magnetic configurations, which allowed at least for some degree of external control.
The possibility of manipulation is, however, much better in mesoscopic systems such as quantum dots.
There, system parameters are widely tunable via applied gate and bias voltages, in addition to external fields. 
This is one motivation for many theoretical and experimental studies of single and double quantum dots coupled to superconducting leads (see Refs.~\onlinecite{de_franceschi_hybrid_2010,martin-rodero_josephson_2011} for an overview). 	For example, Cooper-pair splitting in double quantum dot (DQD) devices has attracted a lot of attention\cite{Schonenberger,Hermann,Heiblum}. The possibility to control the nonlocal entanglement in a Cooper pair splitting device by means of spin-orbit interaction together with inter- as well as intradot Coulomb interaction has been proposed theoretically\cite{Hussein2016}.

Recently, we proposed DQDs as a minimal model to generate all four classes of superconducting pairing in a single device\cite{WeissUnconv} and discussed how triplet pairing influences the electric current in various transport regimes.
The discussed signatures, however, did not distinguish between even- and odd-triplet pairing and both are, in general, generated simultaneously.
To circumvent this problem, we focus in this paper on a single quantum dot hosting one orbital level only.
In this case, the lack of orbital degrees of freedom implies that odd-triplet pairing is the only unconventional paring mechanism that can occur in addition to the conventional even-singlet one; even-triplet and odd-singlet are ruled out by symmetry. It has been shown recently, that odd-triplet correlations might be rigourously detected by a Majorana STM\cite{BjoernMajo}. 
For a full removal of spin symmetry, inhomogeneous magnetic fields, as proposed for DQDs\cite{WeissUnconv}, cannot be employed for a single-level quantum dot.
Instead, we suggest to couple the quantum dot to two ferromagnetic leads with noncollinear magnetization directions.
The feasibility to achieve this has been experimentally demonstrated recently\cite{Kontos}.

A theoretical treatment of quantum dots coupled to superconductors in nonequilibrium situations is complicated by the presence of Coulomb interaction in the quantum dot.
If the Coulomb interaction is weak, it can be treated perturbatively\cite{fazio98,schwab99,clerk00,cuevas01,avishai03,dell-anna08,koerting10}.
To allow for arbitrarily strong Coulomb interaction, one can perform a perturbation expansion in the tunnel coupling between dot and leads\cite{pala07,governale08}.
In the limit of an infinitely-large superconducting gap in the lead, an exact treatment of both the Coulomb interaction and the tunnel coupling between quantum dot and superconducting leads is possible
\cite{governale08,rozhkov00,tanaka07,karrasch08}.
Here, we employ the diagrammatic real-time technique developed for single-\cite{governale08} and double\cite{Eldridge}-quantum dots attached to superconductors, which also allows for calculating shot noise\cite{Braggio}.
While the diagrammatic technique is formulated (and has been applied\cite{David}) for arbitrary values of the superconducting gap $\Delta$ in the lead, we assume, in the following,  the limit of $\Delta \rightarrow \infty$.

The structure of the article is as follows.
First, in Sec.~\ref{sec:model}, we define the model, introduce superconducting order parameters to describe conventional and unconventional correlations, and present the technique to calculate the order parameters as well as the current and current noise.
Then, in Sec.~\ref{sec:intact_spin}, we analyze the case of full spin symmetry. 
In contrast to our previous studies\cite{governale08,Braggio}, we allow for an arbitrary ratio of tunnel-coupling strengths to the superconducting ($\Gamma_S$) and the normal ($\Gamma_N$) leads, instead of relying on either of the limits $\Gamma_S / \Gamma_N \gg 1$ or $\ll 1$.
Because of full spin symmetry, triplet pairing does not occur.
In Sec.~\ref{sec:results_uniaxial}, we treat the case of partially reduced spin symmetry by an applied magnetic field or by involving one ferromagnetic lead.
Finally, Sec.~\ref{sec:asymm} is devoted to the case of full removal of spin symmetry by means of two, noncollinearly-magnetized, ferromagnetic leads.

\section{Model \& Method}
\label{sec:model}
\begin{figure}
	\includegraphics[width=\columnwidth]{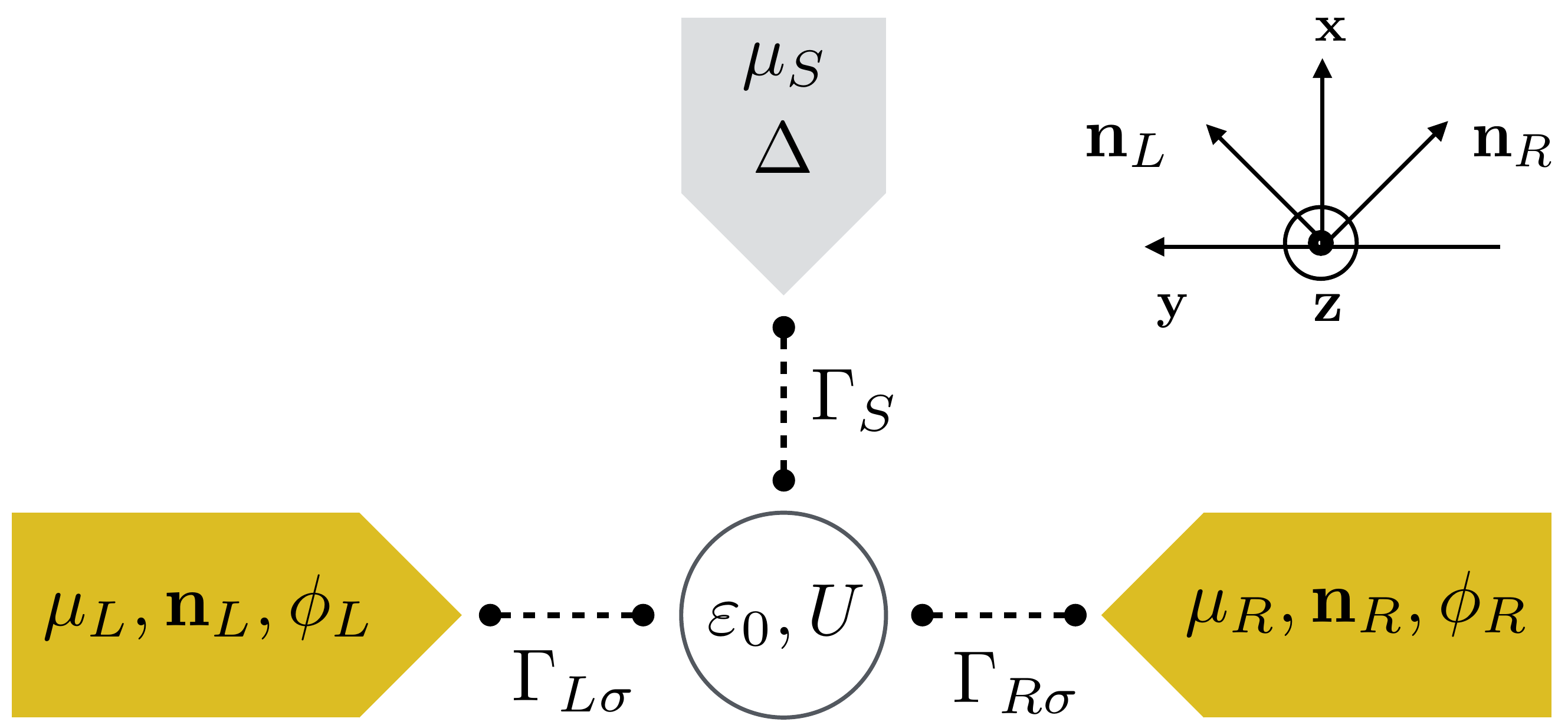}
	\caption{\label{fig:fig1}(Color online) A single-level quantum dot is tunnel coupled to two ferromagnetic leads with coupling strength $\Gamma_\alpha$, $\alpha=L/R$, magnetized in directions ${\bf n}_{\alpha}$ and a conventional superconductor with gap $\Delta$. The magnetization of each ferromagnet encloses an angle $\phi_\alpha$ with the ${\bf x}$-axis. Tunneling between superconductor and quantum dot with coupling strength $\Gamma_S$ gives rise to a finite pairing amplitude on the quantum dot. The underlying coordinate system is shown as well.}
\end{figure}
In order to explore unconventional superconducting correlations in a single-level quantum dot, we use the model depicted in Fig.~\ref{fig:fig1}. A quantum dot containing a single (spinful) orbital is coupled to  a conventional BCS superconductor denoted by the label $S$.
Spin symmetry may be reduced by an external magnetic field and/or coupling the quantum dot to two ferromagnetic leads ($L$ and $R$). 

\subsection{Hamiltonian}

The Hamiltonian is
\begin{equation}
\mathcal{H}=\mathcal{H}_\text{dot}+\sum_{\alpha=L,R,S}\left(\mathcal{H}_\alpha+\mathcal{H}_{\text{tun},\alpha}\right).
\label{Hsys}
\end{equation}
The quantum dot is described by the Hamiltonian
\begin{equation}
	\mathcal{H}_\text{dot}=\sum_{\sigma}\epsilon_0 d_\sigma^\dag d_\sigma 
	+g\mu_B{\boldsymbol B}\cdot{\boldsymbol S}/\hbar+Un_\uparrow n_\downarrow,
\label{eq:proxH}
\end{equation}
with fermionic operators $d_\sigma^\dag (d_\sigma)$ that create (annihilate) electrons on the quantum dot with spin $\sigma$. 
We denote by $n_\sigma=d_\sigma^\dag d_\sigma$ the particle-number operator. 
The single-particle energy $\epsilon_0$ can be tuned by a gate voltage.
The parameter $U$ models the charging energy for double occupation of the dot. 
An externally applied magnetic field is denoted ${\boldsymbol B}$ and the dot spin is ${\boldsymbol S}=(\hbar/2)\sum_{\sigma\sigma'}d^\dag_\sigma {\boldsymbol \sigma}_{\sigma\sigma'}d_{\sigma'}$, with the vector of Pauli matrices ${\boldsymbol \sigma=(\sigma_x,\sigma_y,\sigma_z)}$. 

To treat the superconducting lead, we assume the limit of a large superconducting gap, $\Delta\to \infty$\cite{governale08,rozhkov00,tanaka07,karrasch08}.
This allows us to integrate out the superconductor to arrive at an effective Hamiltonian for the quantum dot, $\mathcal{H}_\text{dot}+H_\text{S}+H_\text{tun,S}\to \mathcal{H}_\text{eff}$ with
\begin{equation}
	\mathcal{H}_\text{eff}=\mathcal{H}_\text{dot}-\frac{\Gamma_S}{2}
	\left(d^\dag_\uparrow d^\dag_\downarrow+d_\downarrow d_\uparrow\right),
\end{equation}
with $\Gamma_S=2\pi |t_S|^2\rho_S$ where $\rho_S$ is the (spin-independent) density of states for electrons in the normal state at the Fermi energy in the superconductor. 
In the absence of a Zeeman field, $\boldsymbol{B}=0$, the spectrum of the effective Hamiltonian is given by $\epsilon_0$ for the single-electron states $|\uparrow\rangle$ and $|\downarrow\rangle$ and $\epsilon_\pm=\delta/2\pm \epsilon_A$ with $2\epsilon_A=\sqrt{\delta^2+\Gamma_S^2}$ and detuning $\delta=2\epsilon_0+U$ for the coherent superposition of an empty and a doubly-occupied dot,  
\begin{align}
	|\pm\rangle&=\frac{1}{\sqrt{2}}\left(\sqrt{1\mp\frac{\delta}{2\epsilon_A}}|0\rangle\mp \sqrt{1\pm \frac{\delta}{2\epsilon_A}}|d\rangle\right) \, ,
\label{eq:pm}
\end{align}  
refered to as Andreev bound states.

The two ferromagnetic leads, $\alpha=L,R$, are described by grand canonical Hamiltonians
\begin{align}
	\mathcal{H}_\alpha&=\sum_{k\tau}\left(\epsilon_{k\alpha\tau}-\mu_\alpha\right)c^\dag_{k\alpha\tau}c_{k\alpha\tau}
\end{align}
where $\mu_\alpha$ is the electrochemical potential of lead $\alpha$, which can be changed by applying bias voltages. 
We assume noninteracting electrons in the leads having energy $\epsilon_{k\alpha\tau}$ and use operators $c_{k\alpha\tau}^\dag (c_{k\alpha\tau})$ to create (annihilate) an electron in lead $\alpha$ with spin $\tau$ and momentum $k$. 
The spin quantization axis can be chosen differently for each lead. 
For the ferromagnets, it is conventient to choose the magnetization directions ${\bf n}_L$ and ${\bf n}_R$, respectively, as natural quantization axes, such that $\tau=\pm$ describe majority and minority spins with density of states $\rho_\pm$ at the Fermi energy.
The degree of spin polarization at the Fermi energy is given by $p_\alpha=(\rho_{\alpha+}-\rho_{\alpha-})/(\rho_{\alpha+}+\rho_{\alpha-})$.
Paramagnetic leads are described by $p_\alpha=0$, whereas $p_\alpha=1$ corresponds to a half-metallic lead hosting majority spins only. 
The hybridization between dot and lead $\alpha$ due to a tunneling amplitude $t_\alpha$ yields a finite line width $\Gamma_{\alpha\pm}=2\pi|t_\alpha|^2\rho_{\alpha\pm}$, which, for  $p_\alpha\neq 0$, is spin dependent.
We define the average line width as $\Gamma_\alpha:=\sum_{\sigma=\pm}\Gamma_{\alpha\sigma}/2$. 

Tunneling between the quantum dot and the leads with tunneling amplitude $t_\alpha$ is modeled by 
\begin{equation}
	\mathcal{H}_{\text{tun},\alpha}=\sum_{k \tau\sigma} \left(t_\alpha c^\dag_{k\alpha\tau} u^\alpha_{\tau\sigma}d_\sigma+h.c.\right),
\end{equation}
where the unitary operator $u^{\alpha}_{\tau\sigma}$ describes the rotation of the dot's spin coordinate system to that of lead $\alpha$.
To be specific, we choose the quantization axis for the dot electrons along the $z$-axis, while ${\bf n}_\alpha$ lie in the $x$-$y$-plane, enclosing an angle $\phi_\alpha$ with the ${\bf x}$-axis, see Fig.~\ref{fig:fig1}. 
This fixes the coordinate axes to be ${\bf e}_x=({\bf n}_R+{\bf n}_L)/|{\bf n}_R+{\bf n}_L|$, ${\bf e}_y=({\bf n}_R-{\bf n}_L)/|{\bf n}_R-{\bf n}_L|$, ${\bf e}_z={\bf e}_x\times {\bf e}_y$ \cite{QDSV_1,QDSV_2}.  
As a result, we get explicitly
\begin{align}
	\mathcal{H}_{\text{tun},\alpha}&=\frac{t_{\alpha}}{\sqrt{2}}\sum_{k}\left[c^\dag_{k
	\alpha+}\left(e^{i\phi_\alpha/2}d_{\up}+e^{-i\phi_\alpha/2}d_{\down}
	\right)\right.\nonumber\\
	&+ \left.c^\dag_{k\alpha-}\left(-e^{i\phi_\alpha/2}d_{\up}+e^{-i\phi_\alpha/2}d_{\down}
	\right)\right]+h.c. 
\end{align}

\subsection{Kinetic equation}

An applied bias voltage between any two leads $\alpha$ and $\beta$ yields a difference $eV=\mu_\alpha-\mu_\beta$ in the electrochemical potentials.
In such a nonequilibrium situation, a net charge current can flow.
The main idea of the diagrammatic technique used here is to analyze the system's dynamics in terms of the reduced density matrix ${\bf P}\equiv \left( P_{\chi_1}^{\chi_2} \right)$ of the dot degrees of freedom that is obtained after integrating out the leads.
Diagonal elements $P_{\chi}^\chi$ describe the nonequilibrium occupation probabilities of the respective states, whereas off-diagonal elements $P_\chi^{\chi'}$ describe coherent superpositions.
The kinetic equation for the reduced density matrix is given by
\begin{eqnarray}
	\frac{d}{dt} P^{\chi_1}_{\chi_2}(t) &=& -i \sum_\chi \left(
	h_{\chi_1\chi}P^{\chi}_{\chi_2} -
	h_{\chi \chi_2} P^{\chi_1}_{\chi} \right) (t) 
	\nonumber \\
	&& +\sum_{\chi_1' \chi_2'} \int_{-\infty}^t dt' W_{\chi_2\chi_2'}^{\chi_1 \chi_1'} (t,t') P_{\chi_1'}^{\chi_2'} (t').  
 	\label{rhodot}
\end{eqnarray}
The first line describes the coherent evolution of the system due to the effective Hamiltonian $\mathcal{H}_\text{eff}$ with matrix elements $h_{\chi\chi'}=\langle\chi|H_{\text {eff}}|\chi'\rangle$.
The second line, involving the irreducible self-energy kernels $W$, describes both coherent and dissipative dynamics due to coupling to the ferromagnets.
In the steady-state limit, the time derivatives of the reduced density matrix elements vanish.
The stationary density matrix element is, then, obtained from
\begin{eqnarray}
\label{stationary}
	0= -i \sum_\chi \left(
	h_{\chi_1\chi}P^{\chi}_{\chi_2} -
	h_{\chi \chi_2} P^{\chi_1}_{\chi} \right)
	+\sum_{\chi_1' \chi_2'} W_{\chi_2\chi_2'}^{\chi_1 \chi_1'} P_{\chi_1'}^{\chi_2'} \, ,  
\end{eqnarray}
where we introduced the generalized transition rates $W_{\chi_2\chi_2'}^{\chi_1 \chi_1'} := \int_{-\infty}^t dt' W_{\chi_2\chi_2'}^{\chi_1 \chi_1'} (t,t')$.

For the basis states $\chi$, it is natural to use $\left\{|0\rangle,|\up\rangle,|\down\rangle,|d\rangle\right\}$. 
In addition to the kinetic equations for the diagonal reduced density matrix elements, we need to include the off-diagonal matrix elements describing the coherence between $|0\rangle$ and $|d\rangle$.
In the regime $\Gamma_S\gg\Gamma_N$, it is advantageous to use the basis $\left\{|\up\rangle,|\down\rangle,|+\rangle,|-\rangle\right\}$ instead since, then, only diagonal matrix elements need to be taken into account and compact analytic formulas can be obtained for various transport quantities\cite{Sothmann, Braggio}.
In the following, however, we will extend those previous studies by making no restriction on the ratio $\Gamma_S/\Gamma_N$.
As a consequence, we need to keep all off-diagonal elements that are induced by the presence of the superconductor and/or the ferromagnets.
All the time, however, we will assume that the tunnel coupling between quantum dot and leads $\alpha=L/R$ is weak such that only generalized transition rates up to first order in $\Gamma_\alpha$ will be included. 
They are computed with the help of diagrammatic rules presented in Ref.~\onlinecite{Sothmann}.

\subsection{Superconducting order parameters}
\label{sec:observables}

Superconducting correlations between two electrons of spin $\sigma$ and $\sigma'$ are quantified by the Green's function
\begin{equation}
	F_{\sigma \sigma'}(t)=\langle\T d_{\sigma}(t)d_{\sigma'}(0)\rangle \, ,
\label{eq:gorkov}
\end{equation}
where $\text{T}$ is the time-ordering operator.
It is an anomalous (Gorkov) Green's function as it involves two annihilation operators and, therefore, probes coherences between particle-number states that differ by two electrons.
The time arguments of the two operators indicate that Eq.~\eqref{eq:gorkov} measures correlations between electrons not only at equal but also at different times.
To construct out of Eq.~\eqref{eq:gorkov} Green's functions that possess definite symmetry in the spin and time arguments, we form proper linear combinations. 
For even-singlet correlations, we define
\begin{equation}
\label{singlet}
	F_e^S(t) = \frac{F_{\uparrow\downarrow}(t) - F_{\downarrow\uparrow}(t)}{2} \, , 
\end{equation}
which is a scalar under rotation in spin space.
Furthermore, odd-triplet correlations are characterized by the three combinations
\begin{subequations}
\begin{align}
	F_o^{T_x}(t) &= \frac{F_{\downarrow\downarrow}(t) - F_{\uparrow\uparrow}(t)}{\sqrt{2}}
\\
	F_o^{T_y}(t) &= -i \frac{F_{\downarrow\downarrow}(t) + F_{\uparrow\uparrow}(t)}{\sqrt{2}}
\\
	F_o^{T_z}(t) &= \frac{F_{\downarrow\uparrow}(t) + F_{\uparrow\downarrow}(t)}{\sqrt{2}} \, ,
\end{align}
\label{triplet}
\end{subequations}
that transform under rotations like the cartesian components of a vector ${\vec F}(t)$.
Since there are only four possibilities to choose $\sigma$ and $\sigma'$ in Eq.~\eqref{eq:gorkov}, we cannot construct any other linear combination that is independent from Eqs.~\eqref{singlet} and \eqref{triplet}. For a single level quantum dot, fixed symmetry in spin space immediately determines the symmetry of the Gorkov Green's function in frequency (time) \cite{foot1}.

The Gorkov Green's functions contain the full information about the time-dependence of the superconducting correlations.
To define an order parameter, we want to characterize the correlations by a single number only.
A natural candidate for such an order parameter is the equal-time correlator.
This works well for even-singlet correlations, which yields the dimensionless order parameter
\begin{equation}
\label{delta_even_relation}
	\Delta_e^S := F_e^S(0) = \langle d_\downarrow d_\uparrow \rangle = P_0^d \, ,
\end{equation}
which is nothing but the amplitude of the coherent superposition of the dot being empty and doubly occupied. 

For odd-frequency correlations, however, $F_o^{T_i}(0), i=x,y,z$ vanishes due to symmetry, and another definition of the order parameter is required.
Following Ref.~\onlinecite{balatsky_even-_1993}, we define the order parameter as the derivative of the Gorkov Green's function at equal time, 
\begin{equation}
	{\boldsymbol \Delta}_o^T := \hbar \frac{d{\vec F}_o(t)}{dt} \bigg|_{t=0} \, ,
\label{delta_odd}
\end{equation}
which has units of energy. Importantly enough, in the small quantum dot studied here, there is no room to construct Green's functions describing even-triplet or odd-singlet correlations.
Throughout this paper, we assume a weak tunnel coupling between quantum dot and leads $\alpha=L/R$, such that only first-order tunnel processes from and to the normal leads need to be taken into account.
This is consistent with discussing the superconducting order parameters to lowest, i.e., zeroth order in $\Gamma_N$ only.
In this limit, we can completely express the odd-triplet order parameters in terms of reduced-density-matrix elements of the quantum dot.
To do so, we first evaluate $\hbar \dot{d}_\sigma(t)=i [\mathcal{H}_\text{eff},d_\sigma](t)$ and, then, plug the result into $\hbar \langle \dot d_\sigma (0) d_{\sigma'}(0) \rangle$, which yields
\begin{subequations}
\begin{align}
 	\hbar \langle \dot d_\uparrow d_\uparrow \rangle &= -\frac{i}{2} (B_x-iB_y) \langle d_\downarrow d_\uparrow \rangle
	+\frac{i}{2}\Gamma_{\text{S}} \langle d_\downarrow^\dagger d_\uparrow \rangle
\\
	\hbar \langle \dot d_\uparrow d_\downarrow \rangle &= -i\epsilon \langle d_\uparrow d_\downarrow \rangle
	-\frac{i}{2}B_z \langle d_\uparrow d_\downarrow \rangle
	+\frac{i}{2}\Gamma_{\text{S}} \langle d_\downarrow^\dagger d_\downarrow \rangle
\\
	\hbar \langle \dot d_\downarrow d_\uparrow \rangle &= -i\epsilon \langle d_\downarrow d_\uparrow \rangle
	+\frac{i}{2}B_z \langle d_\downarrow d_\uparrow \rangle
	-\frac{i}{2}\Gamma_{\text{S}} \langle d_\uparrow^\dagger d_\uparrow \rangle
\\
 	\hbar \langle \dot d_\downarrow d_\downarrow \rangle &= -\frac{i}{2} (B_x+iB_y) \langle d_\uparrow d_\downarrow \rangle
	-\frac{i}{2\hbar}\Gamma_{\text{S}} \langle d_\uparrow^\dagger d_\downarrow \rangle
\end{align}
\label{tderiv}
\end{subequations}
Combining this results with Eqs.~\eqref{triplet} and \eqref{delta_odd} leads to the very compact result
\begin{equation}
	\label{delta_odd_relation}
	{\boldsymbol \Delta}_o^T = \frac{i}{\sqrt{2}} {\vec B} \Delta_e^S - \frac{i}{\sqrt{2}\hbar}\Gamma_S {\vec S} \, .
\end{equation}
It indicates that odd-triplet correlations in a single-level quantum dot are either associated with even-singlet correlations in the presence of a Zeeman field or with a finite spin polarization in the presence to a tunnel coupling to a superconducting lead.
Thereby, it is irrelevant whether the finite spin polarization is induced by a Zeeman field or by non-equilibrium spin accumlation due to voltage-biased ferromagnetic leads. Note that the Coulomb interaction strength $U$ does not enter explicitely Eqs.~\eqref{tderiv}. However, it is taken into account when calculating expectation values, e.g. $\Delta_e^S$ and $\vec{S}$ in Eq.~\eqref{delta_odd_relation}, with respect to the full Hamiltonian \eqref{Hsys}.  

We remark that Eqs.~\eqref{delta_even_relation} and \eqref{delta_odd_relation} are valid both in equilibrium and in non-equilibrium.
To compute thereof order parameters, one needs to find $\rho_0^d$ and $\vec S$ with the help of the kinetic quations of the reduced density matrix.

\subsection{Charge current}

Voltages applied to the leads drive currents through the quantum dot.
The current flowing from the normal lead $\alpha=L,R$ into the quantum dot can be calculated from 
\begin{equation}
	I_\alpha=\frac{e}{2 \hbar} \text{Tr}\left( {\bf W}^{I_\alpha}{\bf P} \right) \, .
	\label{cur}
\end{equation}
Here, the stationary density matrix element $P_{\chi_1}^{\chi_2}$ has to be determined from the kinetic equation in the stationary limit, Eq.~\eqref{stationary}, and the matrix elements $\left(W^{I_\alpha}\right)_{\chi\chi_2}^{\chi\chi_1}$ of ${\bf W}^{I_\alpha}$ are the current rates. 
They are given by the generalized transition rates multiplied with the number of electrons transfered from lead $\alpha$ to the dot.
Rules for an explicit calculation are given in Ref.~\onlinecite{Sothmann}. The current $I_S$ into the superconductor, referred to as Andreev current, follows from current conservation,
\begin{equation}
	I_S=I_L+I_R \, .
\end{equation}

\subsection{Current noise and cross correlations}

In addition to current, we also study zero-frequency current fluctuations, given by
\begin{align}
\label{noise}
	S_{\alpha\beta}&=\lim_{\omega\rightarrow 0} \int\limits_{-\infty}^\infty dt \langle \delta I_\alpha(t)\delta I_\beta(0)+\delta I_\beta(0)\delta I_\alpha(t)\rangle e^{-i\omega t} \, ,
\end{align}
with $\delta I_\alpha=I_\alpha-\langle I_\alpha\rangle$.
For $\alpha = \beta$, the correlations Eqs.~\eqref{noise} describe the shot noise power, for $\alpha \neq \beta$ cross correlations between lead $\alpha$ and $\beta$. We have ensured that 
the definition, Eq.~\eqref{noise}, reproduces results of an equivalent study based on derivatives of the cumulant generating function \cite{Braggio}. Following Ref.~\onlinecite{Braun_noise}, this can be expressed diagrammatically as
\begin{align}
	S_{\alpha\beta}=\frac{e^2}{2 \hbar}\lim_{\omega\rightarrow 0} \text{Tr} & 
	\left[\left( {\bf W}^{I_\alpha I_\beta} + {\bf W}^{I_\beta I_\alpha} \right.\right.
\nonumber \\
	& \left. + {\bf W}^{I_\alpha}\left[{\bf \Pi}^{-1}_0(\omega)-{\bf W}\right]^{-1}{\bf W}^{I_\beta} \right.
\nonumber \\
	& \left.\left. + {\bf W}^{I_\beta}\left[{\bf \Pi}^{-1}_0(-\omega)-{\bf W}\right]^{-1}{\bf W}^{I_\alpha}\right){\bf P} \right] \, .
\label{corr_nl}
\end{align}
In this expression the kernels ${\bf W}^{I_\alpha I_\beta}$ are obtained from $W_{\chi_2\chi_2'}^{\chi_1 \chi_1'}$ in Eq.~\eqref{rhodot} by replacing two tunneling vertices by current vertices, the one earlier in time by $I_\beta$ and the one later in time by $I_\alpha$.
To lowest order in $\Gamma_N$, this term is diagonal in the {\it lead indices}, ${\bf W}^{I_\alpha I_\beta}  \sim \delta_{\alpha \beta}$. 
The free propagation of the system between the successive current measurements, that enters in Eq.~\eqref{corr_nl}, is given by
\begin{equation}
	\left(\Pi_0\right)^{\chi_1,\chi_1'}_{\chi_2,\chi_2'}(\omega)=\frac{i\delta_{\chi_1,\chi_1'}\delta_{\chi_2,\chi_2'}}{\epsilon_{\chi_2}-\epsilon_{\chi_1}-\hbar \omega +i 0^+}.
\end{equation}
The term $\hbar \omega$ in the denominator prevents the appearance of divergencies for $\chi_1=\chi_2$ for any finite $\omega$.
In the limit $\omega \rightarrow 0$, i.e., the zero-frequency noise, the sum of the terms occuring in Eq.~\eqref{corr_nl} is well defined.

\section{Results}
\label{sec:results}
The kinetic equation Eq.~\eqref{rhodot} is solved numerically taking into account matrix elements $\chi_1,\chi_2\in\{|0\rangle,|{\uparrow}\rangle,|{\downarrow}\rangle,|d\rangle\}$, i.e. we obtain all occupation probabilities together with the overlaps between the respective states to first order in $\Gamma_N$. Within this weak-tunneling approximation, i.e. $\Gamma_N\ll k_BT$, the current as well as the noise are calculated with the help of Eq.~\eqref{cur} and Eq.~\eqref{corr_nl}.  
For the proximitized dot, $\Gamma_S\neq 0$, in the presence of FM leads with $\phi_\alpha\neq 0$, we have eight independent nonvanishing elements of $P_{\chi}^{\chi'}$: four occupation probabilities and four off-diagonal elements are different from zero due to the presence of the superconductor, $P_0^d=(P_d^0)^*\neq0$ as well as due to the FM leads $P_\uparrow^\downarrow=(P_\downarrow^\uparrow)^*\neq 0$. The tensor of irreducible kernels  ${\bf W}$ 
is of dimension $8\times 8$ in our case. 
We rearrange Eq.~\eqref{rhodot} in the stationary limit such that ${\bf 0}={\bf \tilde{W}} {\bf P}$ and solve for ${\bf P}$ by calculating the nullspace of ${\bf \tilde{W}}$, a matrix that combines both ${\bf W}$ and the first line of Eq.~\eqref{rhodot} involving the effective Hamiltonian.

A subtlety arises from the intrinsic inconsistency of the master equation when expanded to first order in $\Gamma_N$. If the contribution of the dissipative kernels in Eq.~\eqref{rhodot} is much smaller than the line describing coherent evolution, entries of the density matrix may not be of order $\Gamma_N$ but higher order contributions are mixed in. These next-to-leading order effects, for instance the possibility to have finite currents within the Coulomb-blockaded region (e.g. due to cotunneling), scale at least with $\Gamma_N^2$.  Within our approach we have carefully monitored that these artefacts do not have a visible effect in the numerical results.

Recently, several analytic approximations have been put forward. The regime $\Gamma_S,\Gamma_N\ll k_BT$ around the resonance $\delta\approx 0$, where only first-order tunneling processes to either the superconducting or the normal leads needs to be taken into account, was explored in Ref.~\onlinecite{pala07}. 
Within the regime, $\Gamma_N\ll \Gamma_S$, where the dot spectrum is dominated by the formation of Andreev bound states, it is sufficient to take into account diagonal elements of the density matrix for the calculation of the current\cite{Sothmann,Braggio}. They correspond to single occupancy or occupancy of the respective states of Eq.~\eqref{eq:pm}. This approximation is justified as long as the first line in Eq.~\eqref{rhodot} is of order $\Gamma_S\gg\Gamma_N$. Consequently, the regime $\Gamma_S\ll\Gamma_N$ or $\Gamma_S\sim\Gamma_N$ need a different treatment. Overlap matrix elements $P_-^+$ become important, when approaching $\Gamma_S\sim\Gamma_N$. Within this work we focus on transport properties as current, noise and cross correlators in the regime $\Gamma_N\ll\Gamma_S$. For this subgap regime it is possible to resolve the avoided crossings between Andreev bound states most clearly. However, in the opposite regime $\Gamma_S\ll\Gamma_N$, our results are not changed qualitatively.

Three different scenarios are distinguished in the following: (A) the spin symmetry is intact. (B) We include a finite magnetic field along a specific axis that breaks $SU(2)$ symmetry, however a residual $U(1)$ spin symmetry is still kept. In (A) and (B) bias voltages are chosen such that $\mu_S=0, \mu_L=\mu_R$. Within scenario (C), we study the regime with no spin symmetry at all, i.e.  $(p_\alpha\neq 0,\phi_L\neq \phi_R)$ in a nonequilibrium situation $\mu_L\neq \mu_R$. We discuss the Andreev current as well as induced conventional and unconventional order parameters together with shot noise data.  

\subsection{Full spin symmetry}

\label{sec:intact_spin}

This section presents results for scenario (A).  In the absence of magnetic fields, i.e. $p_\alpha=0$ in each of the leads and ${\bf B}=0$, the $SU(2)$ symmetry is intact. The total spin of electrons on the dot and in the paramagnetic leads is conserved. The bias is chosen as $\mu_L=\mu_R=\mu$ for left and right lead, whereas $\mu_S=0$. Units of energy are fixed by the Coulomb interaction $U$ and temperature is set to $k_BT=U/100$.  

\subsubsection{Current}
\begin{figure}[ht!]
\includegraphics[width=\columnwidth]{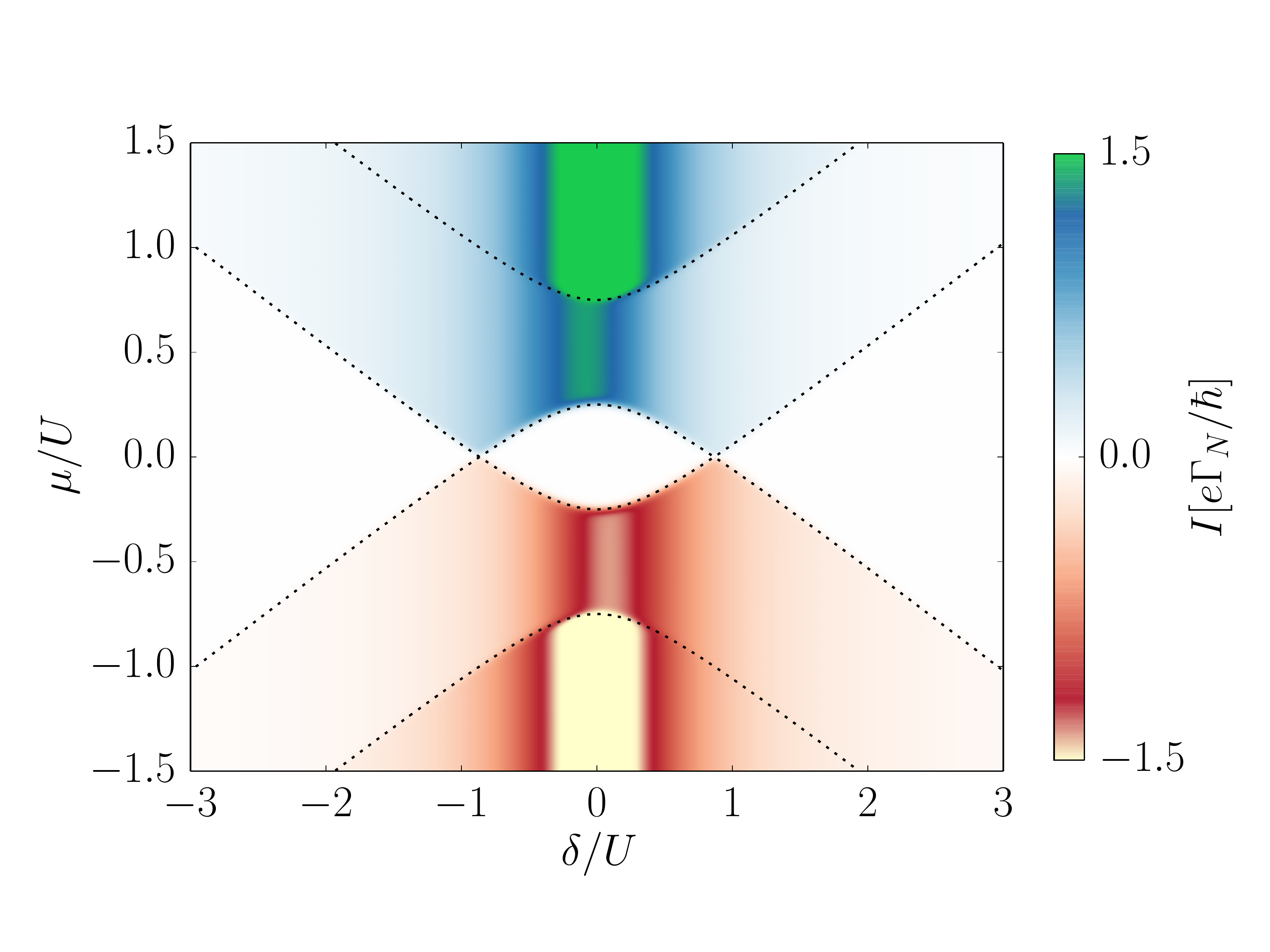}
\caption{\label{fig:fig2}(Color online) Andreev current as a function of gate voltage $\delta$ and  bias voltage $\mu$. The coupling to the SC is $\Gamma_S=U/2$,  the coupling to the paramagnetic  leads is $\Gamma_N=U/1000$ and temperature is $k_BT=U/100$.}
\end{figure}

In Fig.~\ref{fig:fig2} the Andreev current into the superconductor is shown as a function of gate- and bias voltage $(\delta,\mu)$, respectively.  We have chosen a strong coupling of the QD to the superconductor $\Gamma_S=U/2$ to achieve a strong proximity effect.
With respect to the applied bias voltage we distinguish three regimes: (i) low bias $|\mu|<\epsilon_{-}-\epsilon_0$, (ii) intermediate bias $\epsilon_- -\epsilon_0<|\mu|<\epsilon_{+}-\epsilon_0$ and (iii) high bias $|\mu|>\epsilon_{+}-\epsilon_0$. 
The boundaries between these regimes are given by the Andreev addition energies depicted as dashed lines in the figure. In case (i) the dot is mostly occupied with a single electron and around $\delta=0$, a pronounced Coulomb blockade area is developed, i.e. transport is blocked. 
Tuning $\mu>0$ into regime (ii) provides enough energy to allow for a finite current flow into the SC. In that case, a second electron enters the dot and an enhanced probability for Andreev reflection results in a finite current, as two electrons are able to enter the SC as a Cooper pair.  Of course, the opposite process, for $\mu<0$, is possible as well: Cooper pairs are expelled from the superconductor, are split by the charging energy on the dot, and leave the dot towards different or the same lead. 
In general, there is the symmetry $I(\mu,\delta)=-I(-\mu,-\delta)$.
Within the high-bias regime (iii) the Andreev current becomes insensitive to the sign of $\delta$, i.e. the stronger symmetry condition $I(-\delta)=I(\delta)$ holds.

\subsubsection{Shot noise and cross corelations}
\begin{figure}[ht!]
\includegraphics[width=\columnwidth]{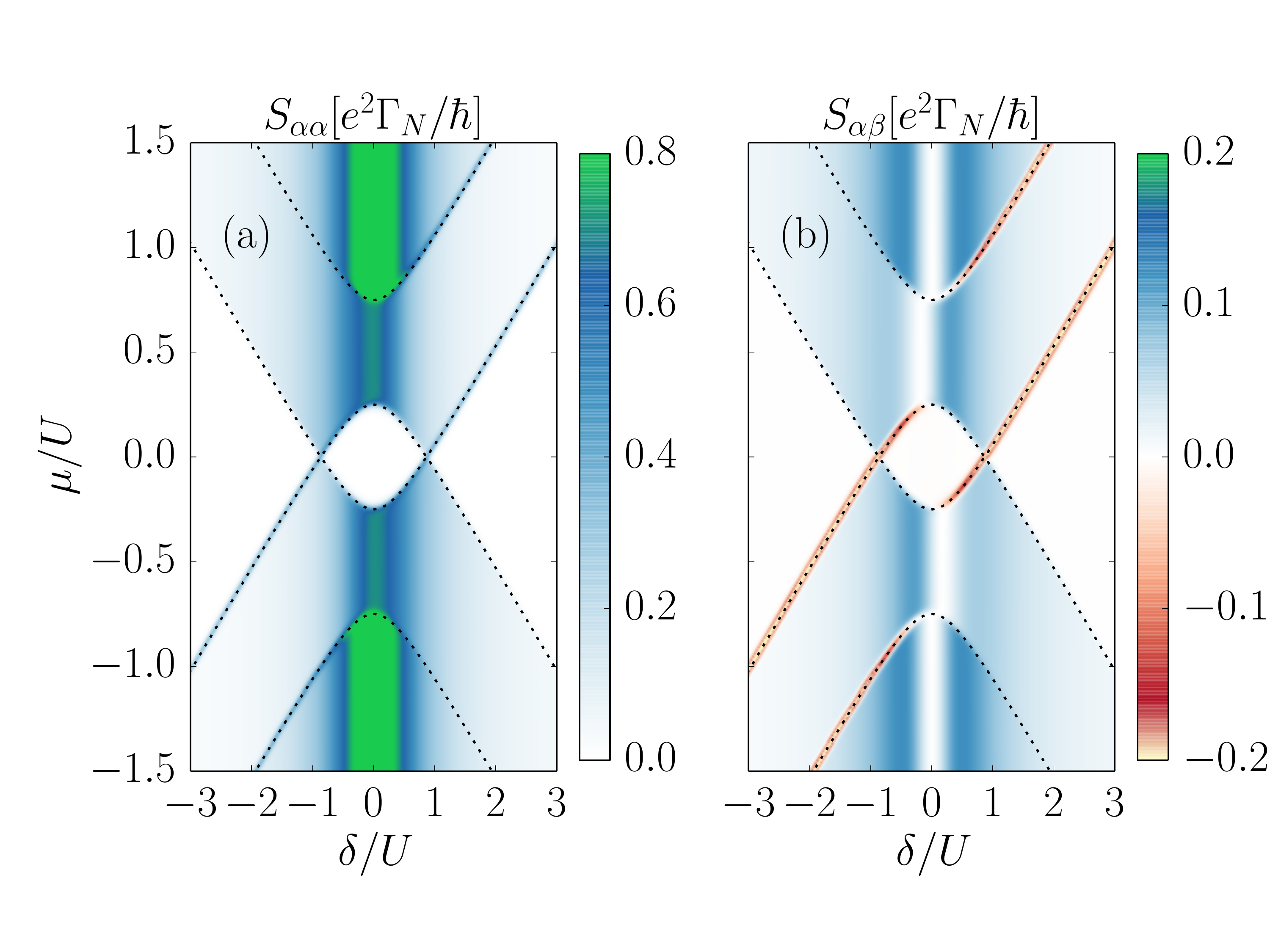}
\caption{\label{fig:fig3}(Color online) Andreev shot noise for (a) same lead (left/or right) and (b) cross-correlations between left and right lead. 
Other parameters are as in Fig.~\ref{fig:fig2}.}
\end{figure}

In Fig.~\ref{fig:fig3}(a), we present data for the shot noise $S_{\alpha\alpha}, \alpha=L \text{ or }R$ of the Andreev current as a function of $(\delta,\mu)$. 
In the low bias regime, the shot noise vanishes as the current does. 
Within regime (ii) and (iii), finite and positive shot noise is present. 
Similar as for the current, there is a main peak of the shot noise around the resonance $\delta=0$.
The main contribution to $S_{\alpha\alpha}$ stems from the first line in Eq.~\eqref{corr_nl}.
In contrast to the current, however, there is an enhancement of the shot noise at the Andreev addition energies in the upper right and lower left part of Fig.~\ref{fig:fig3}(a). 

The cross-correlations $S_{\alpha\beta}, \alpha\neq\beta$ are depicted in Fig.~\ref{fig:fig3}(b). 
There are two noteable features in the figure: first the vanishing of the cross correlations for $\delta\approx 0$ and, second, the appearance of negative cross correlations.  
Since $W^{I_\alpha I_\beta}=W^{I_\beta I_\alpha}=0$ for $\alpha \neq \beta$ in the weak-coupling limit, the only contribution to $S_{\alpha\beta}$ is due to the second and third line in Eq.~\eqref{corr_nl}. For $\delta\approx 0$ these are strongly suppressed and cross correlations vanish. 
Along the Andreev addition-energy lines in the upper right and lower left part of Fig.~\ref{fig:fig3}(b), we obtain $S_{\alpha\beta}<0$, indicating that the transport channels from the left and the right lead block each other\cite{Buettiker_cross,Martin_1992}.

\subsubsection{Even-singlet order parameter}
\begin{figure}[ht!]
\includegraphics[width=\columnwidth]{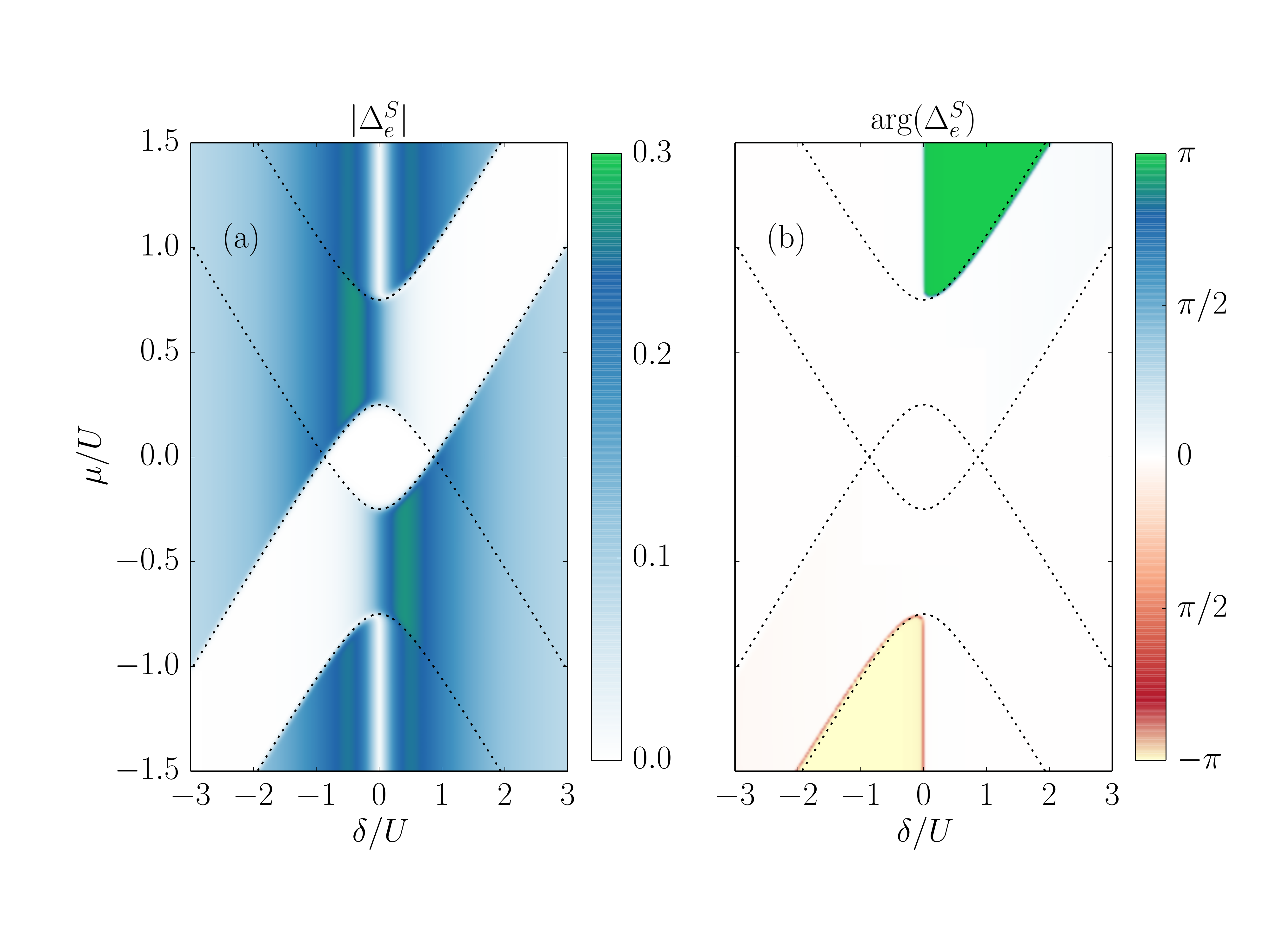}
\caption{\label{fig:fig4}(Color online) (a) Modulus and (b) phase of the induced even-singlet order parameter $\Delta_e^S$ as a function of gate and bias voltage. Other parameter are as in Fig.~\ref{fig:fig2}.}
\end{figure}

A finite coupling of the SC to the quantum dot induces a finite amplitude $\Delta_e^S$ for even-singlet pairing.
It is induced by the proximity effect, i.e., the coupling to the superconducting lead.
In the Hamiltonian $H_S$ for the superconducting lead, Eq.~\eqref{Hsys}, we chose a gauge in which the phase of the superconducting order parameter is zero.
This does not mean, however, that the induced order parameter $\Delta_e^S$ has to have the same phase. 
Since we are dealing with a nonequilibrium situation, $\Delta_e^S$ is complex and its phase contains, in general, nontrivial information.
Therefore, we show in Fig.~\ref{fig:fig4} both the modulus (a) and the phase (b) of $\Delta_e^S$. 
Areas with finite $|\Delta_e^S|$ are present in the intermediate bias regime for $\delta \mu<0$.
At high bias, the order parameter is suppressed around $\delta=0$.
This dip in the modulus of the order parameter as a function of $\delta$ is accompanied with a crossover of the phase from $0$ to $\pi$.
In Fig.~\ref{fig:fig4}, this crossover is rather sharp since we chose $\Gamma_N/\Gamma_S \ll 1$.
With increasing $\Gamma_N$, the crossover becomes smoother.

It has been shown in Refs.~\onlinecite{pala07,governale08} that in the limit of a large superconducting gap in the lead, $\Delta\rightarrow \infty$, the current into the superconductor is directly related to the even-singlet order parameter.
For $\Delta$ real and positive, we get $I_S =(e/\hbar) \Gamma_S \, \text{Im} \, \Delta_e^S$, i.e., the Andreev current in Fig.~\ref{fig:fig2} displays, up to a constant prefactor the imaginary part of the even-singlet order parameter. 
The modulus of the even-singlet order parameter, shown in Fig.~\ref{fig:fig4}(a) looks different because it is, for the chosen parameters, dominated by its real part.

We summarize our findings for scenario (A). In the presence of $SU(2)$ spin symmetry, a conventional spin singlet order parameter exists on the dot.
In a nonequilibrium situation a $0-\pi$-transition can be observed. 
The noise looks qualitatively similar to the current but the cross correlations display a suppression around zero detuning.

\subsection{Residual $U(1)$ spin symmetry}
\label{sec:results_uniaxial}

We now consider the reduction of the $SU(2)$ spin symmetry to $U(1)$, achieved either of the two following ways.
For scenario (B1), the proximitized QD is coupled to FM leads with parallel magnetizations ${\bf n}_L\parallel {\bf n}_R$ between left and right side, accompanied by finite polarizations $p_\alpha\neq 0$, see Fig.~\ref{fig:fig1}.  
The second scenario (B2) is realized, when spin symmetry is broken by a local static magnetic field ${\bf B}=B\hat{{\bf x}}$ applied to the QD.
In this situation, the leads are chosen to be paramagnetic metals ($p_\alpha=0$).

\subsubsection{Scenario B1}
\begin{figure}[ht!]
\includegraphics[width=\columnwidth]{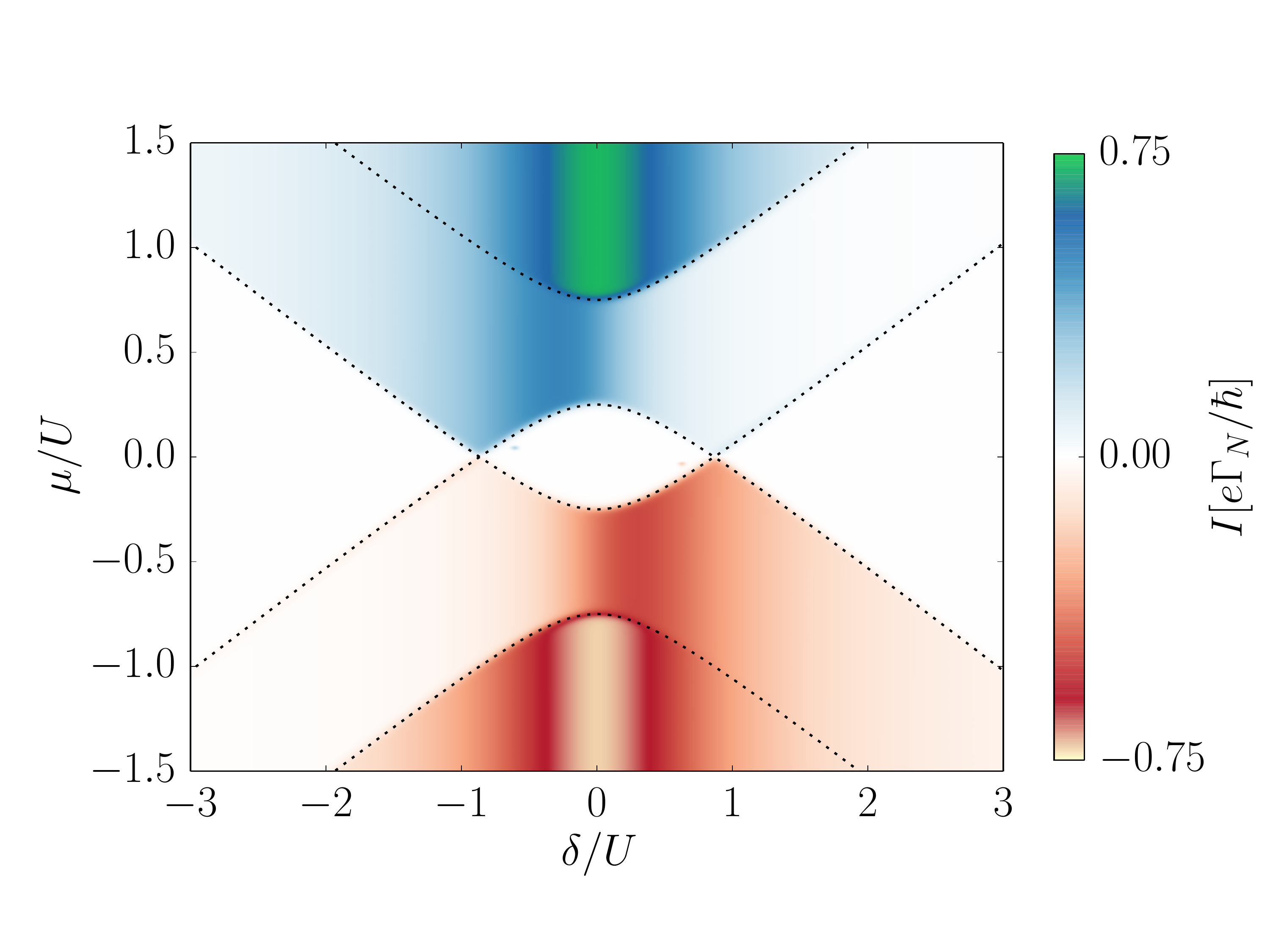}
\caption{\label{fig:fig5}(Color online) Andreev current for scenario (B1), i.e. the proximitized QD is tunnel coupled to ferromagnetic leads. Polarizations are chosen symmetrically $p_\alpha=0.8,\phi_\alpha=0$. Other parameters are as in Fig.~\ref{fig:fig2}.}
\end{figure}

In Fig.~\ref{fig:fig5} we depict $I_S(\delta,\mu)$ for (B1) where the parameters for the FM leads are $\phi_\alpha=0,p_\alpha=0.8, \alpha=L/R$. Temperature and tunnel couplings are chosen as in Fig.~\ref{fig:fig2}. 
As above, the three regimes of low-, intermediate- and large-bias voltage show up. Note that the magnitude of the current is reduced due to the coupling of the QD to ferromagnets. 
Especially within the intermediate bias regime, a clear asymmetry between positive and negative $\delta$ is visible, however the symmetry $I(-\delta,-\mu)=I(\delta,\mu)$ is always intact.

\begin{figure}[ht!]
\includegraphics[width=\columnwidth]{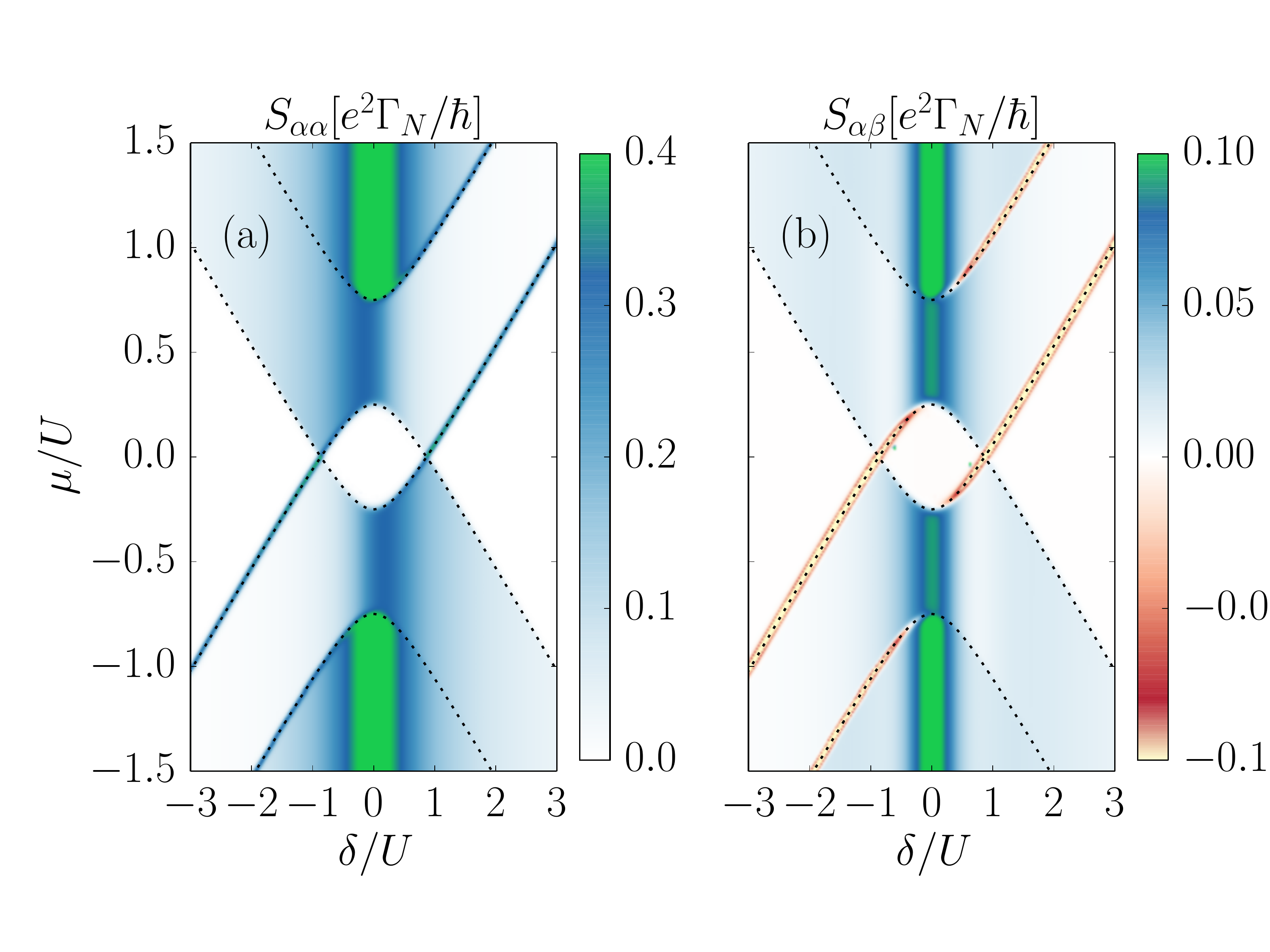}
\caption{\label{fig:fig6}(Color online) (Color online) Andreev shot noise of scenario (B1) for (a) same lead and (b) cross-correlations between left and right lead. 
Other parameters are as in Fig.~\ref{fig:fig5}.}
\end{figure}

Shot noise $S_{\alpha\alpha}$ and cross correlations $S_{\alpha\beta}$ are presented in Fig.~\ref{fig:fig6}(a) and (b), respectively. 
Similar as for the current, the finite spin polarization leads to a reduction of the shot noise and cross correlations.
Many features discussed for scenario (A) hold qualitatively for scenario (B1) as well.
An important difference is that the cross correlations are now not suppressed anymore around zero detuning $\delta\approx 0$.

\begin{figure}[ht!]
\includegraphics[width=\columnwidth]{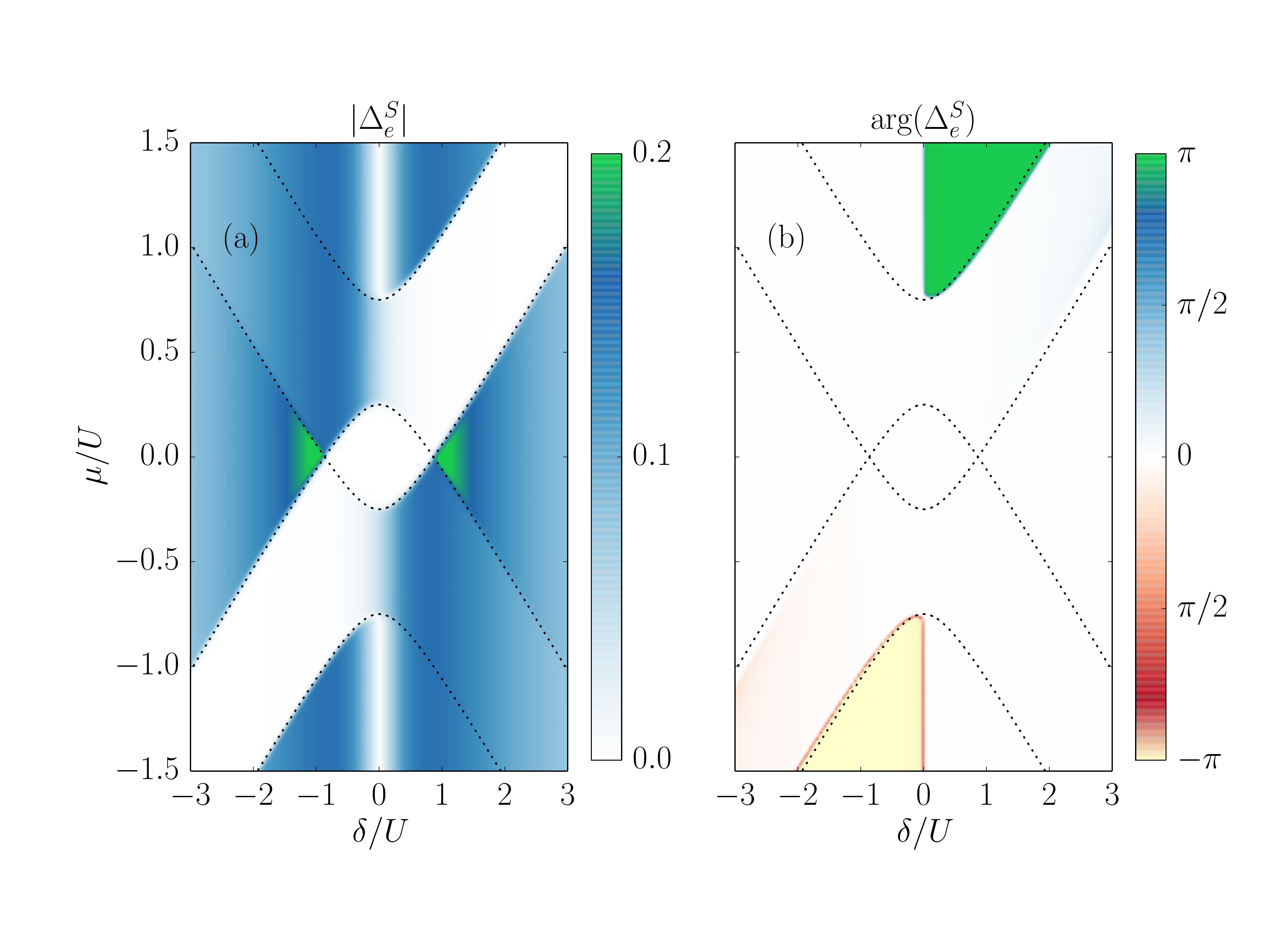}
\caption{\label{fig:fig7}(Color online) (a) Modulus and (b) phase of the induced even-singlet order in scenario (B1) for the same parameters as in Fig.~\ref{fig:fig5}.}
\end{figure}

In Fig.~\ref{fig:fig7}, we show how the presence of the FM leads modify the even-singlet order parameter.
The modulus of $\Delta_e^S$ is, in general, suppressed in magnitude.
In the regime where the QD is most probable singly occupied, however, $|\Delta_e^S|$ is enhanced. 
The phase of $\Delta_e^S$ shows, again a crossover from $0\to\pi$ as $\delta$ goes through zero.

\begin{figure}[ht!]
\includegraphics[width=\columnwidth]{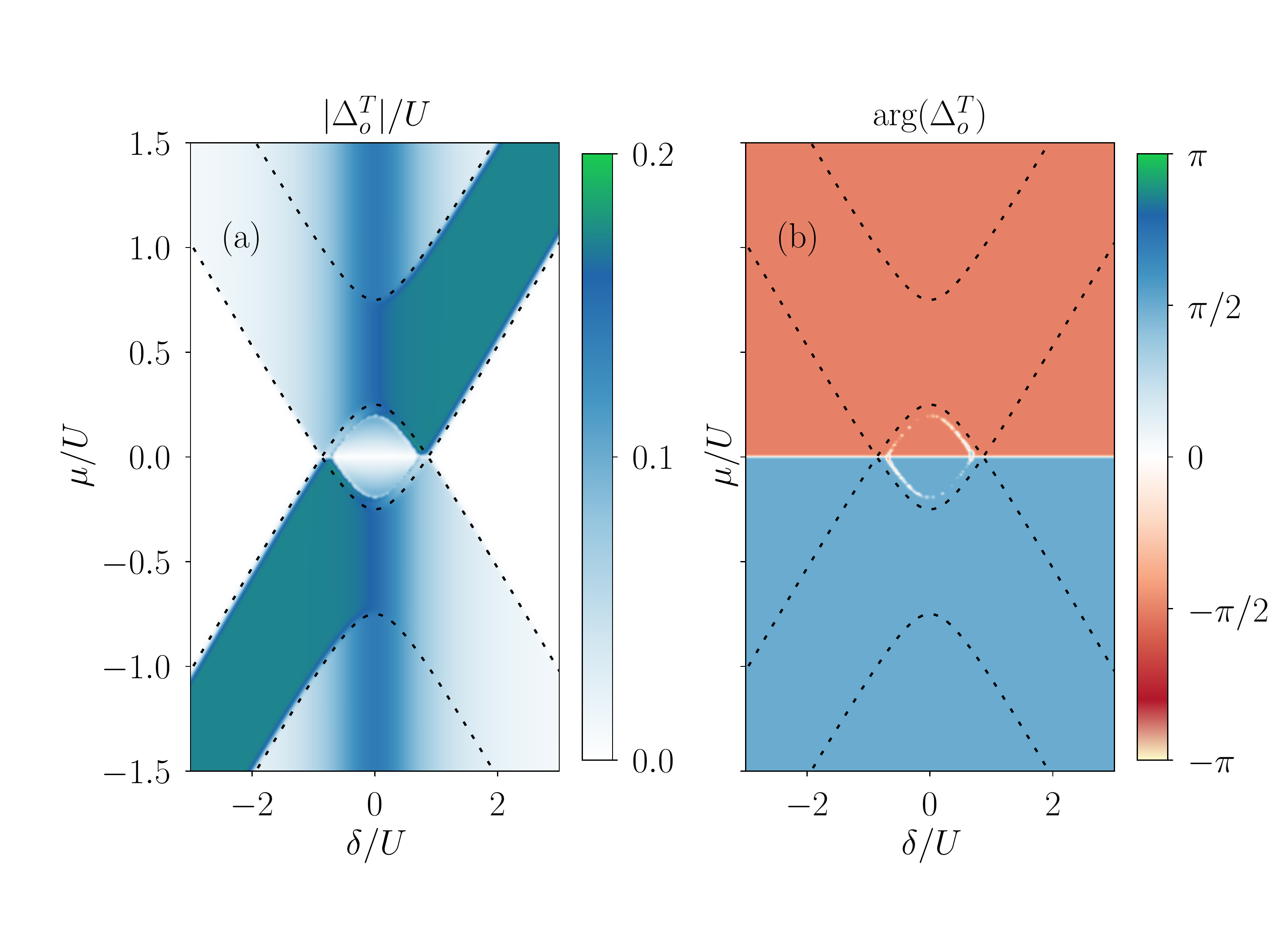}
\caption{\label{fig:fig8}(Color online) Odd-triplet order parameter in scenario (B1). The FM leads yield a finite spin accumulation on the QD, which results in finite $\Delta_o^T$ in the direction of the induced exchange magnetic field. Other parameters are as in Fig.~\ref{fig:fig5}.}
\end{figure}

The most important effect of the FM leads is that now odd-triplet order parameters are induced as well.
According to Eq.~(\ref{delta_odd_relation}), the finite odd-triplet correlation is related to an accumulated dot spin $\mathbf S$ (since there is no magnetic field considered in scenario (B1)).
From Eq.~(\ref{delta_odd_relation}), we conclude that $\mathbf \Delta_o^T$ points along the axis of the residual $U(1)$ symmetry.
The modulus of this component is shown in Fig.~\ref{fig:fig8}(a).
All other components vanishing here. 
Since $\mathbf \Delta_o^T$ is purely imaginary in scenario (B1), the phase is either $\pi/2$ or $-\pi/2$, see Fig.~\ref{fig:fig8}(b). As compared to the even-singlet order parameter, the phase of the odd-triplet one is shifted by $\pi/2$, in accordance with Eq.~\eqref{delta_odd_relation}.

\subsubsection{Scenario B2}
\begin{figure}[ht!]
\includegraphics[width=\columnwidth]{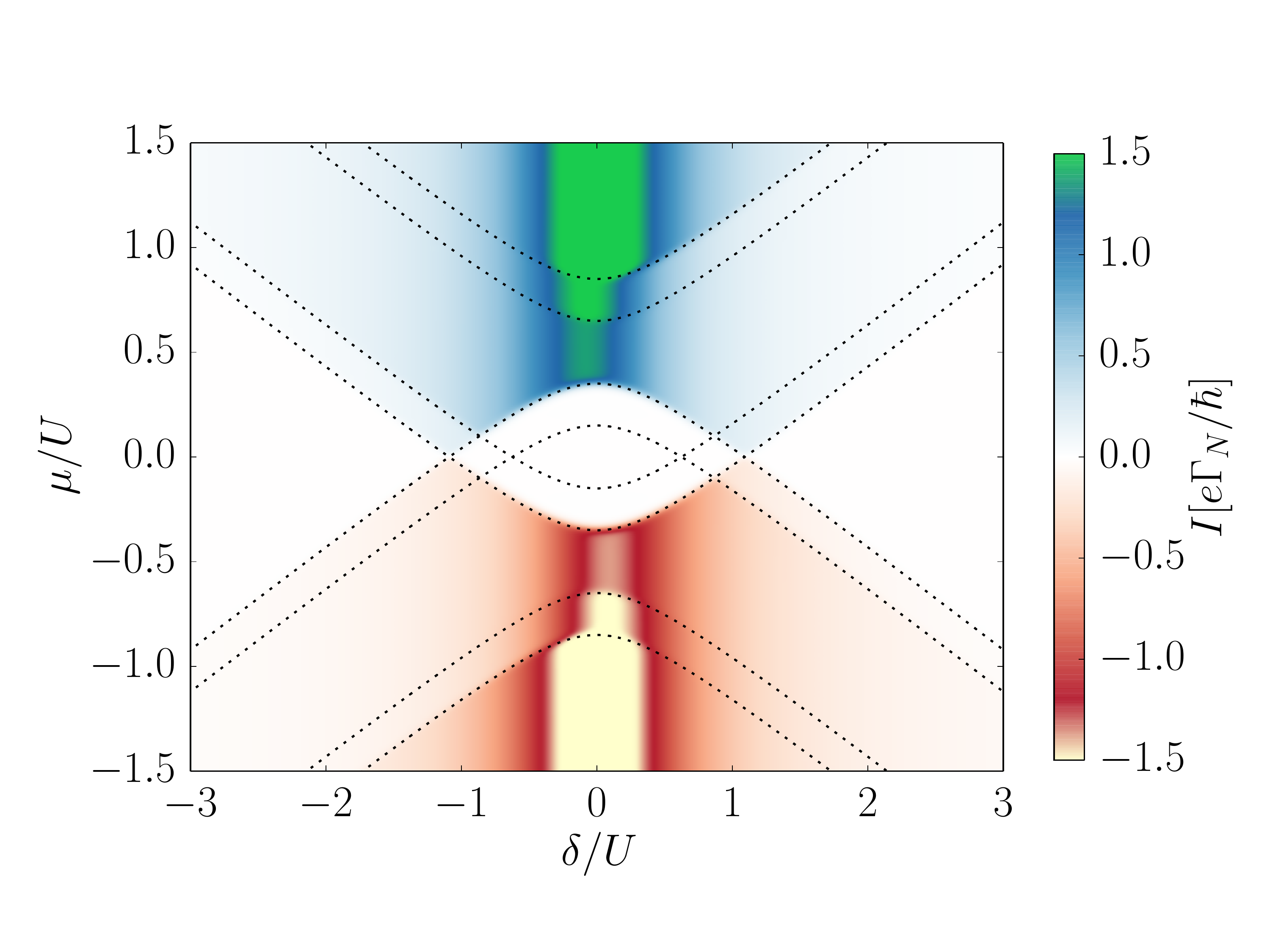}
\caption{\label{fig:fig9}(Color online) Andreev current for scenario (B2). The leads are assumed to be paramagnetic metals $p_\alpha=0$. 
A local magnetic fields $B_x=U/5$, lifts the spin degeneracy on the QD. Other parameters are as in Fig.~\ref{fig:fig2}.}
\end{figure}

In scenario (B2), a local magnetic field applied to the QD is responsible for the reduction of the spin symmetry from $SU(2)$ to $U(1)$, while the leads are paramagnetic, $p_\alpha=0$.
In Fig.~\ref{fig:fig9}, we present the Andreev current $I_S(\delta,\mu)$ as function of detuning and bias voltage.
The Zeeman field acting on the dot enters the Andreev-addition energies.
This generates an additional substructure visible in Fig.~\ref{fig:fig9}.

\begin{figure}[ht!]
\includegraphics[width=\columnwidth]{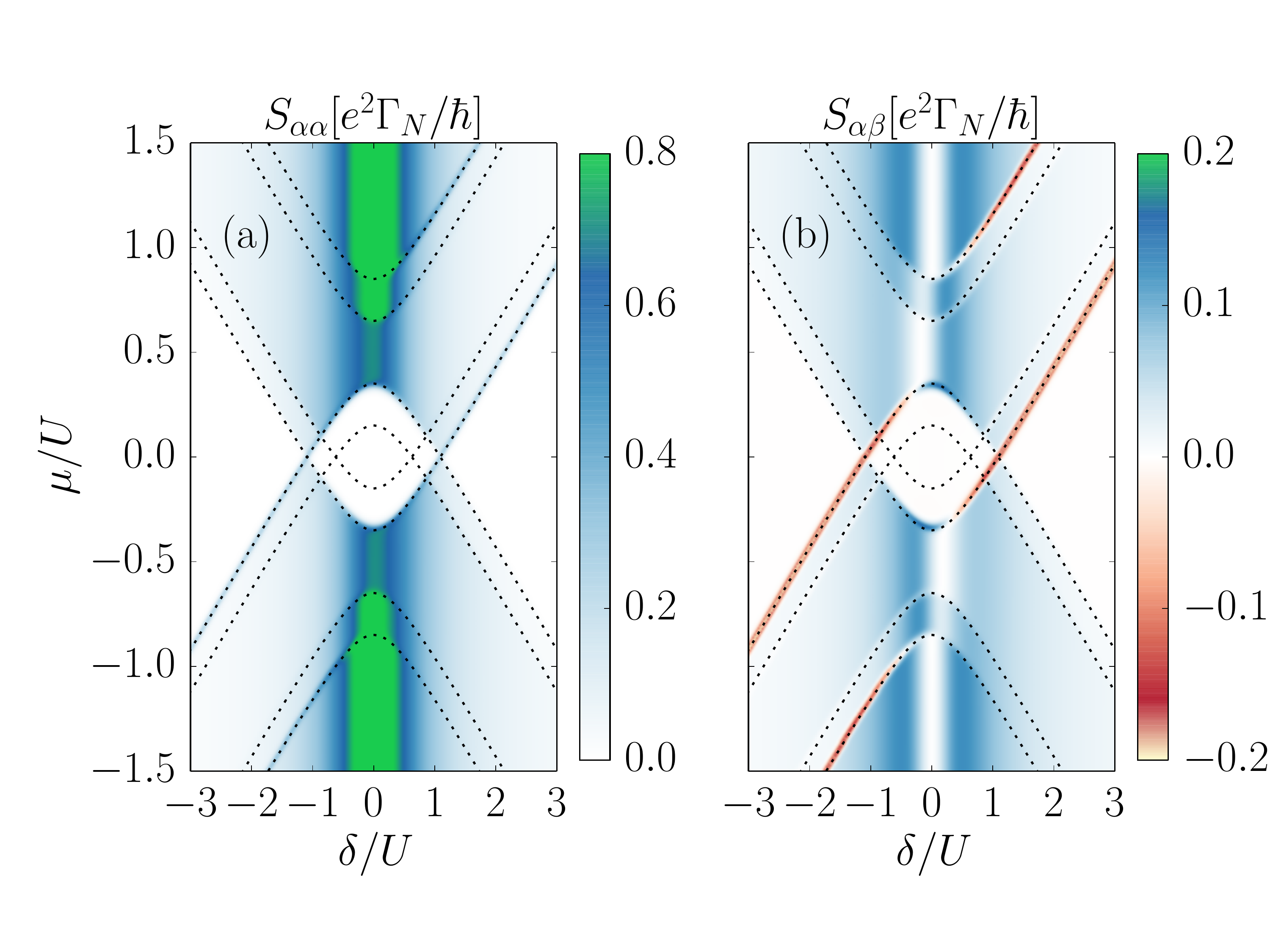}
\caption{\label{fig:fig10}(Color online) Andreev shot noise for scenario (B2), panel (a) depicts the shot noise and in panel (b) the cross correlators are shown. Parameters are as in Fig.~\ref{fig:fig9}. }
\end{figure}

The shot noise $S_{\alpha\alpha}(\delta,\mu)$ is shown in Fig.~\ref{fig:fig10}(a). 
Again, we find a maximum around the resonance $\delta=0$ and a substructure of the Andreev-addition energies, similar as for the current. 
There is an enhancement of the shot noise along two of the Andreev-addition energies in the upper right and lower left.
The cross correlations, see Fig.~\ref{fig:fig10}(b), display a suppression around zero detuning, similar to scenario (A) and in contrast to (B1).

\begin{figure}[ht!]
\includegraphics[width=\columnwidth]{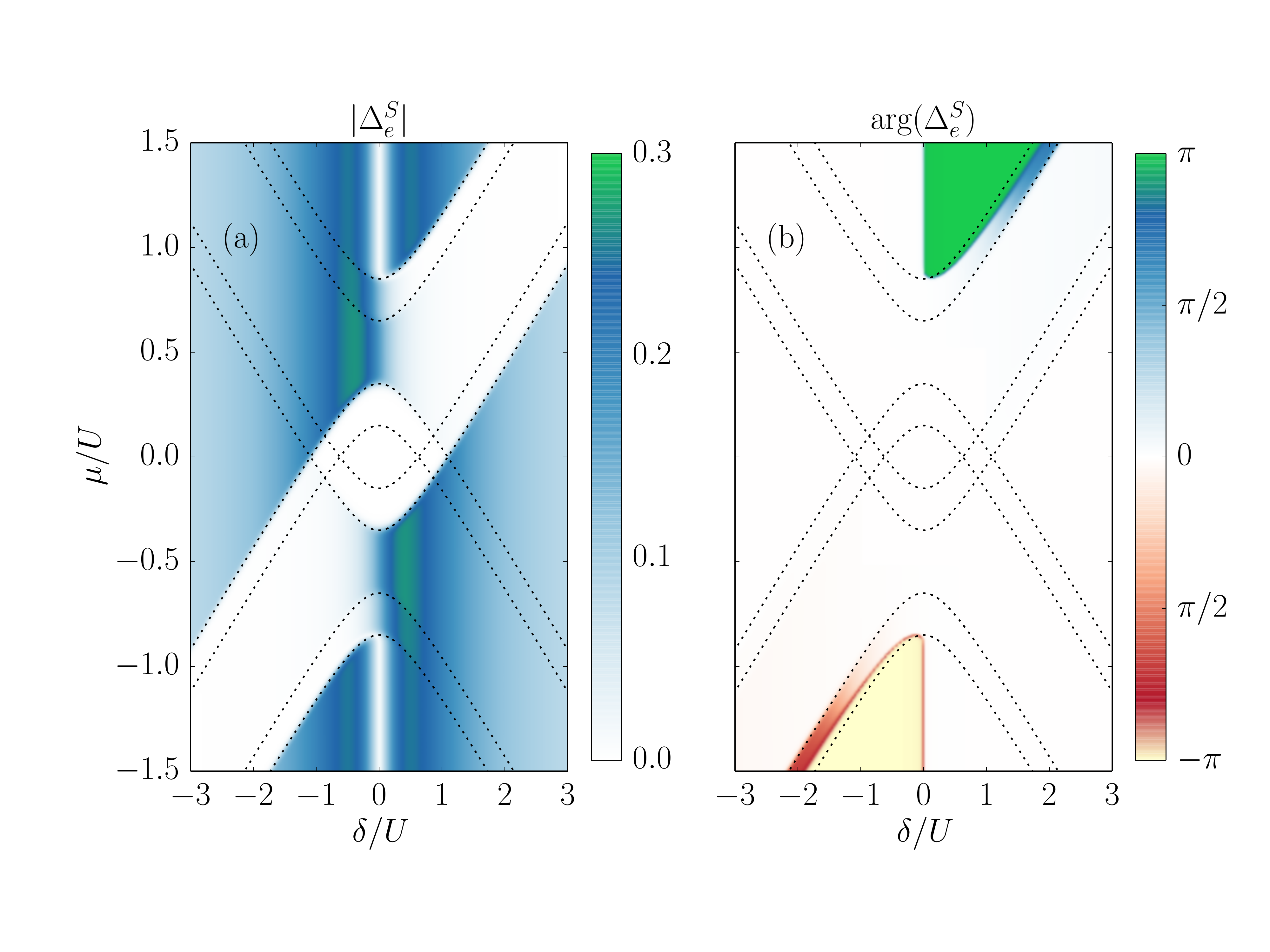}
\caption{\label{fig:fig11}(Color online)  (a) Modulus and (b) phase of the induced even-singlet order parameter scenario B2 for parameters as in Fig.~\ref{fig:fig9}}
\end{figure}

The main impact of the Zeeman field on the induced even-singlet order parameter is the additional substructure in the Andreev-addition energies, see Fig.~\ref{fig:fig11}.
As a consequence, there is now a finite region between the split Andreev-addition in which the transition from phase $0$ to phase $\pi$ becomes smoother (see upper right and the lower left part of Fig.~\ref{fig:fig11}(b)).
The suppression of the even-singlet order parameter around zero detuning is, however, similar as in case (A) and (B1).

\begin{figure}[ht!]
\includegraphics[width=\columnwidth]{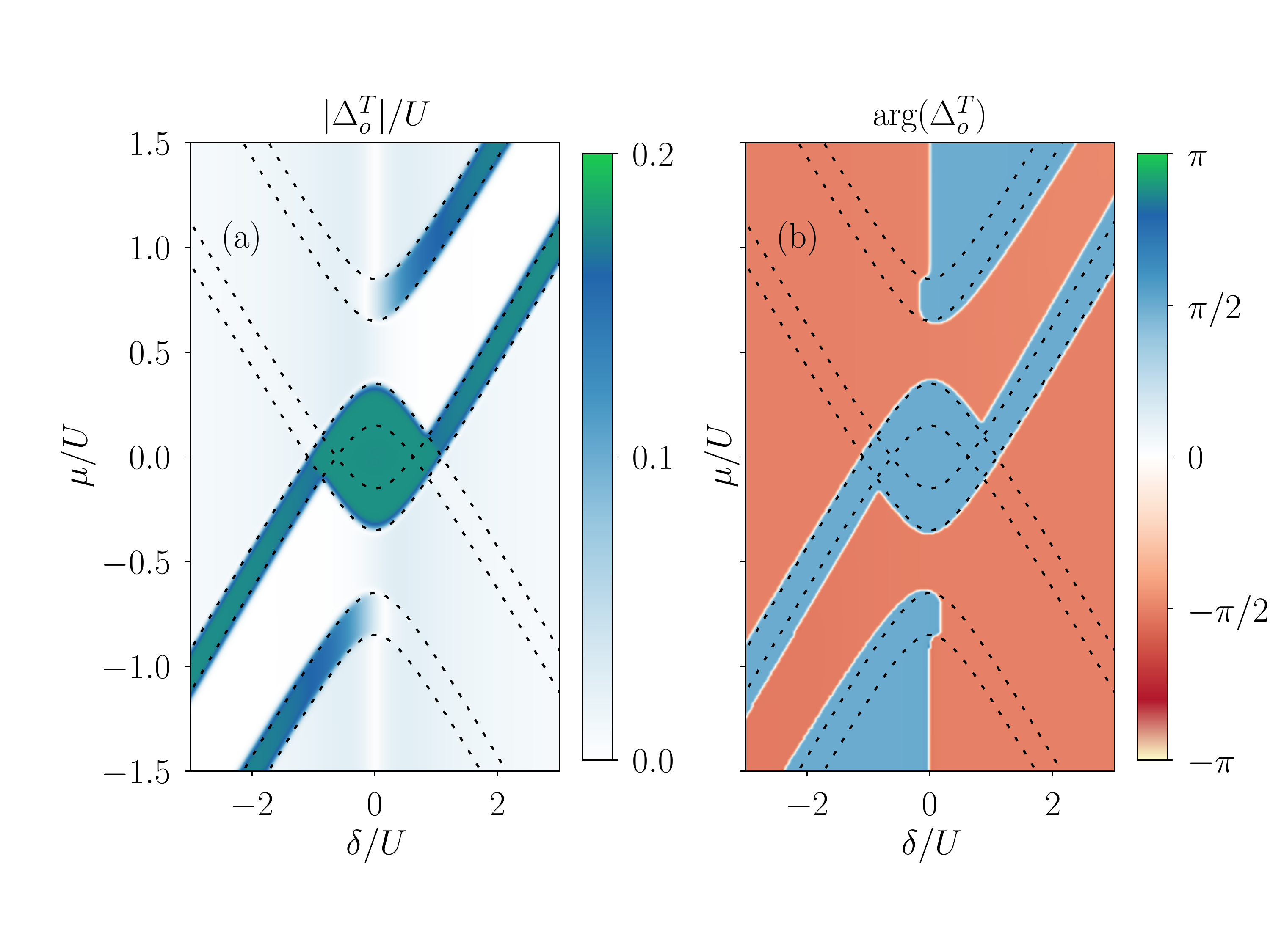}
\caption{\label{fig:fig12}(Color online) Odd triplet order parameter scenario (B2). Parameters are chosen as in Fig.~\ref{fig:fig9}}
\end{figure}

We now turn to the induced odd-triplet order parameter.
It is finite due to the combination of a finite Zeeman field and the presence of even-singlet correlations, see first term in Eq.~\eqref{delta_odd_relation}, and, in addition due to spin accumulation, according to the second term of Eq.~\eqref{delta_odd_relation} as well.
In Fig.~\ref{fig:fig12} modulus (a) and phase (b) of $\Delta_o^{T_x}$, i.e., the component of the odd-triplet order parameter along the direction of the magnetic field, are shown as a function of gate and bias voltage. 
Also here, the phase of the odd-triplet order parameter is shifted by $\pi/2$, see above and Eq.~\eqref{delta_odd_relation}.
The comparison of Fig.~\ref{fig:fig12} with Fig.~\ref{fig:fig11} reveals the interesting result that, for the chosen parameters, odd-triplet correlations do most prominently  appear in regions in which even-singlet correlations are suppressed: in the Coulomb-blockade region in the middle of Fig.~\ref{fig:fig12} and in the small stripes between the split Andreev-addition energies in the upper right and lower left part.
These are regions in which spin accumulation is pronounced.
The very fact that superconducting correlations also show up in the Coulomb-blockade regime underlines the notion of superconducting pairing of electrons at different times, since the probability to have both electron in the quantum at the same time is exponentially suppressed due to charging energy.

To summarize the main feature of scenario (B), a reduction of the spin symmetry, either by an external magnetic field or by FM  leads induces finite odd-frequency order parameters ${\bf \Delta}^T_o$ onto the quantum dot. Triplet order parameters appear in the direction of the applied magnetic or magnetization direction. 
Other components are excluded by the residual spin symmetry.

\subsection{Complete removal of rotational symmetry}

\label{sec:asymm}
\begin{figure}[t]
\includegraphics[width=\columnwidth]{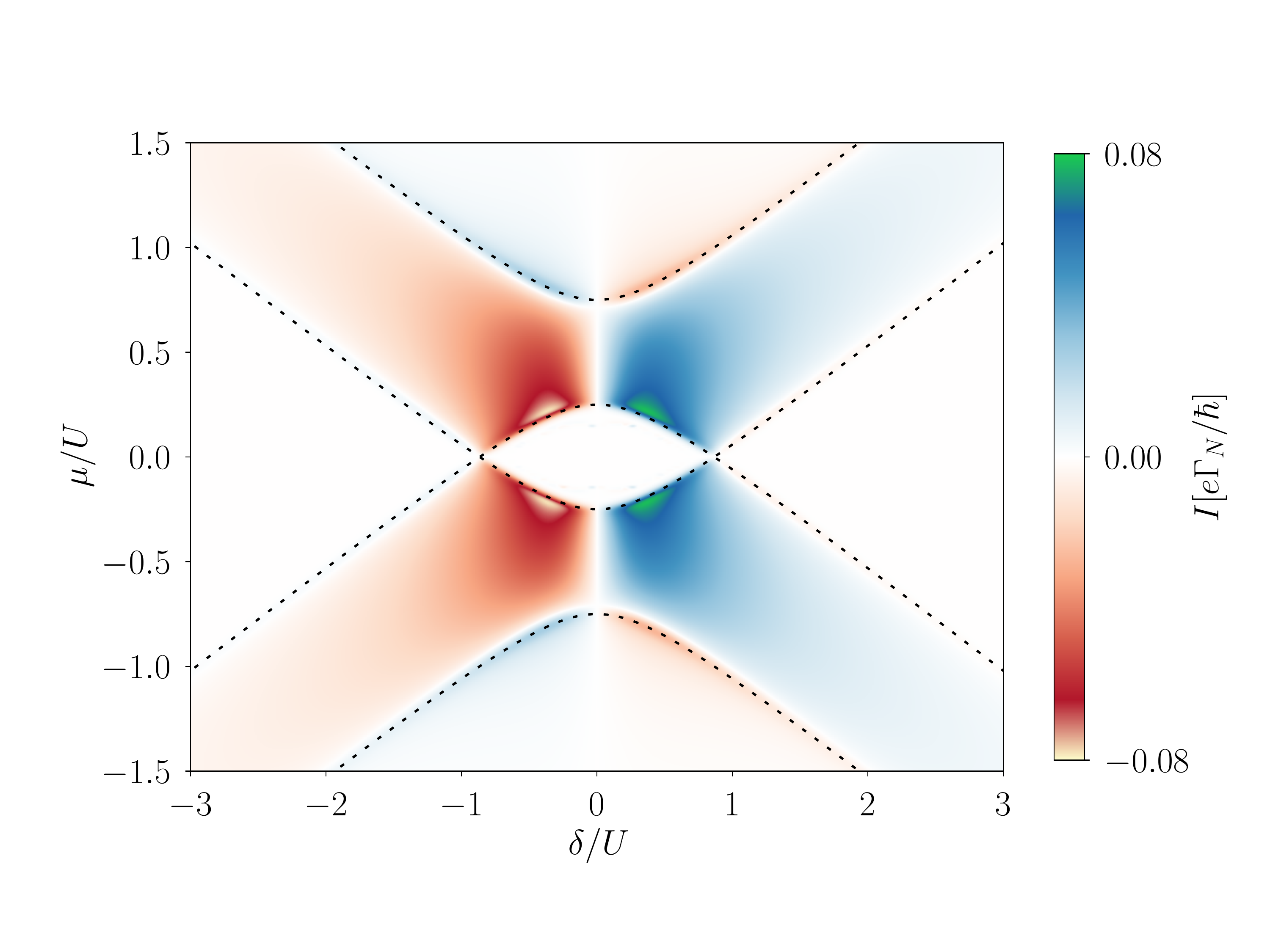}
\caption{\label{fig:fig13}(Color online) Andreev current for scenario (C), here realized with FM leads, finite polarization $p=0.8$, $\phi_L-\phi_R=\pi/2$ and $\mu_L=-\mu_R$. Other parameters are as in Fig.~\ref{fig:fig2}.}
\end{figure}

In scenario (C), spin symmetry is fully broken by two FM leads $p_\alpha\neq 0$ with non-collinear magnetization directions, $\phi_L\neq \phi_R$ that are put at different chemical potentials.
To be specific, we choose $\mu_L=-\mu_R$, relative to $\mu_S=0$ and $\phi_L-\phi_R=\pi/2$.
Choosing different chemical potentials for the two ferromagnets is important for a full removal of spin symmetry. 
Two FM leads with $\phi_L\neq \phi_R$ but $\mu_L=\mu_R$ can be effectively described as one lead with some effective magnetization direction, i.e., the induced superconducting correlation are, in this case, the same as in scenario (B1).

The Andreev current is shown in Fig.~\ref{fig:fig13}.
It now looks, as a consequence of the biasing scheme $\mu_L=-\mu_R$, very different from all the previously discusses cases. 
It is possible to switch between $I_S>0$ and $I_S<0$, i.e. electrons entering or leaving the SC lead, by changing the gate voltage.
The vanishing Andreev current at $\delta=0$ is a consequence of $\mu_L=-\mu_R$ and $\Gamma_L= \Gamma_R$.

\begin{figure}[t]
\includegraphics[width=\columnwidth]{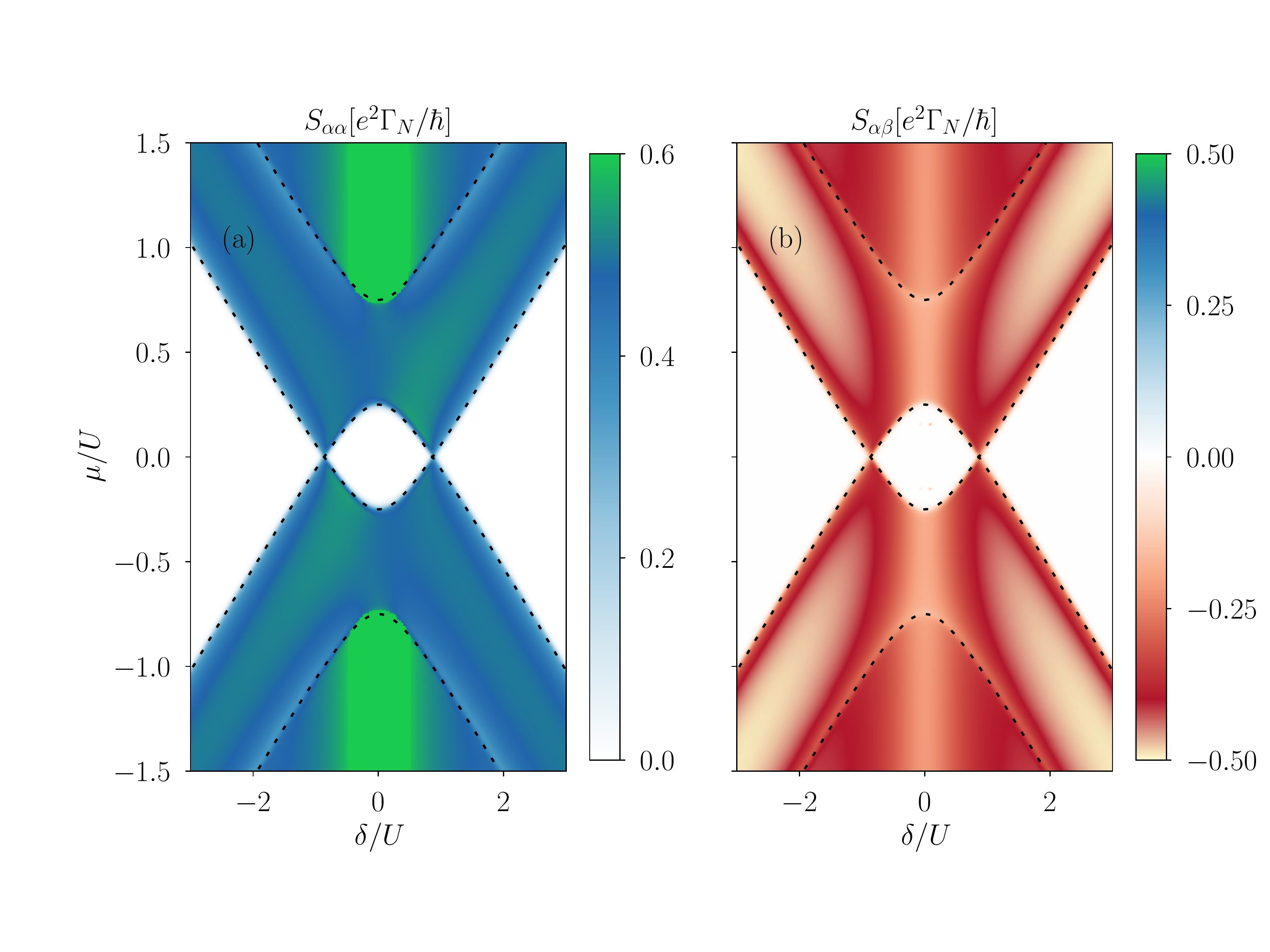}
\caption{\label{fig:fig14}(Color online) (a) Shot noise and (b) cross correlations for scenario (C) for parameters as in Fig.~\ref{fig:fig13}. }
\end{figure}

The corresponding shot noise and cross correlations are shown in Fig.~\ref{fig:fig14}.
We find for all voltages negative cross correlations, $S_{\alpha\beta}<0$ while the shot noise is always positive.

\begin{figure}[t]
\includegraphics[width=\columnwidth]{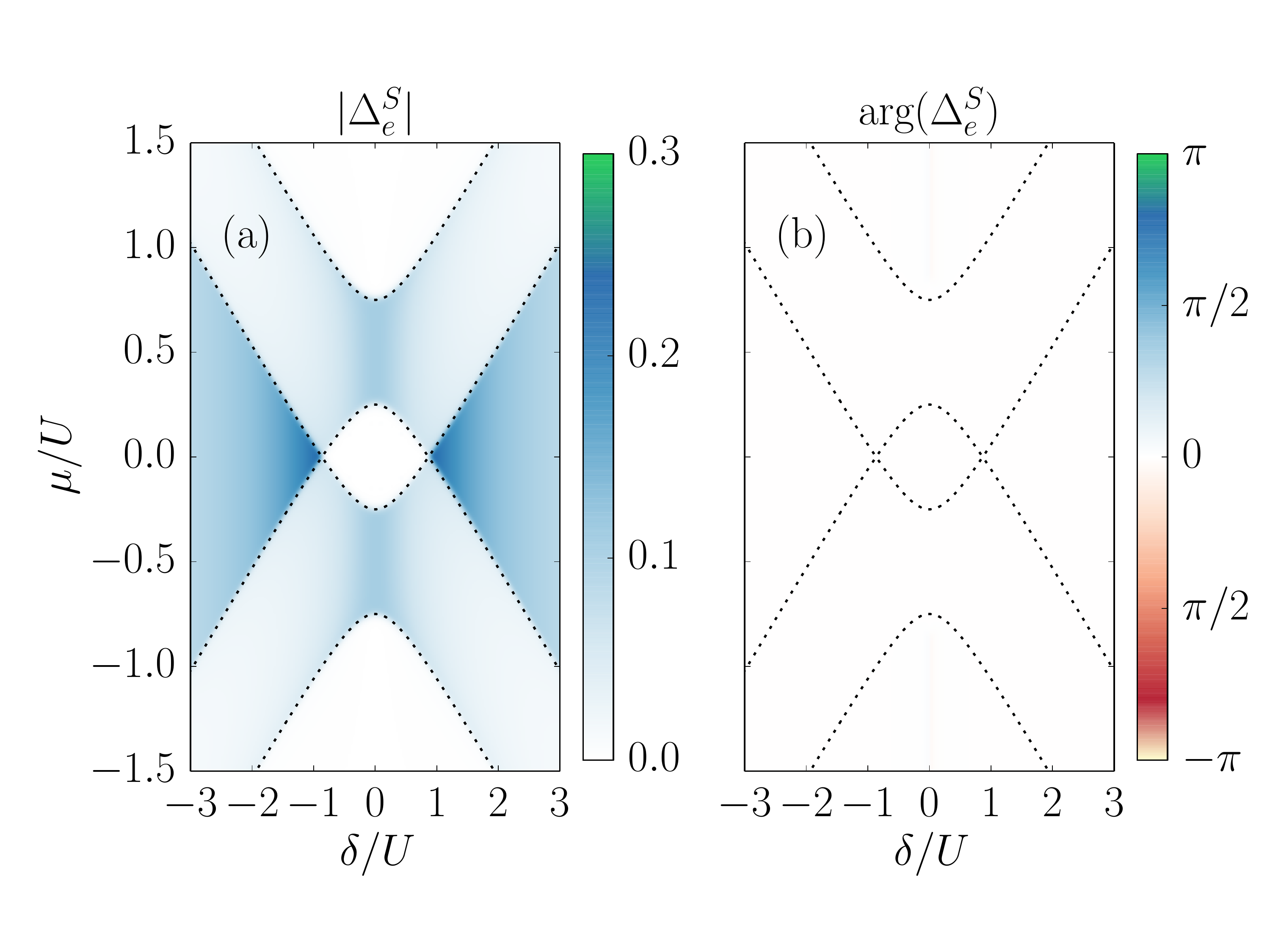}
\caption{\label{fig:fig15}(Color online) (a) Modulus and (b) phase of the even-singlet order parameter for scenario (C) for parameters as in Fig.~\ref{fig:fig13}. }
\end{figure}

The even-singlet order parameter is shown in Fig.~\ref{fig:fig15}.
The condition $\mu_L=-\mu_R$ together with $\Gamma_L=\Gamma_R$ results in $\Delta_e^S$ being real, positive and symmetric with respect to $\delta \rightarrow -\delta$.
The imaginary part is suppressed due to cancellation of different tunneling processes to the left and right side\cite{Sothmann}. 

\begin{figure}[t]
\includegraphics[width=\columnwidth]{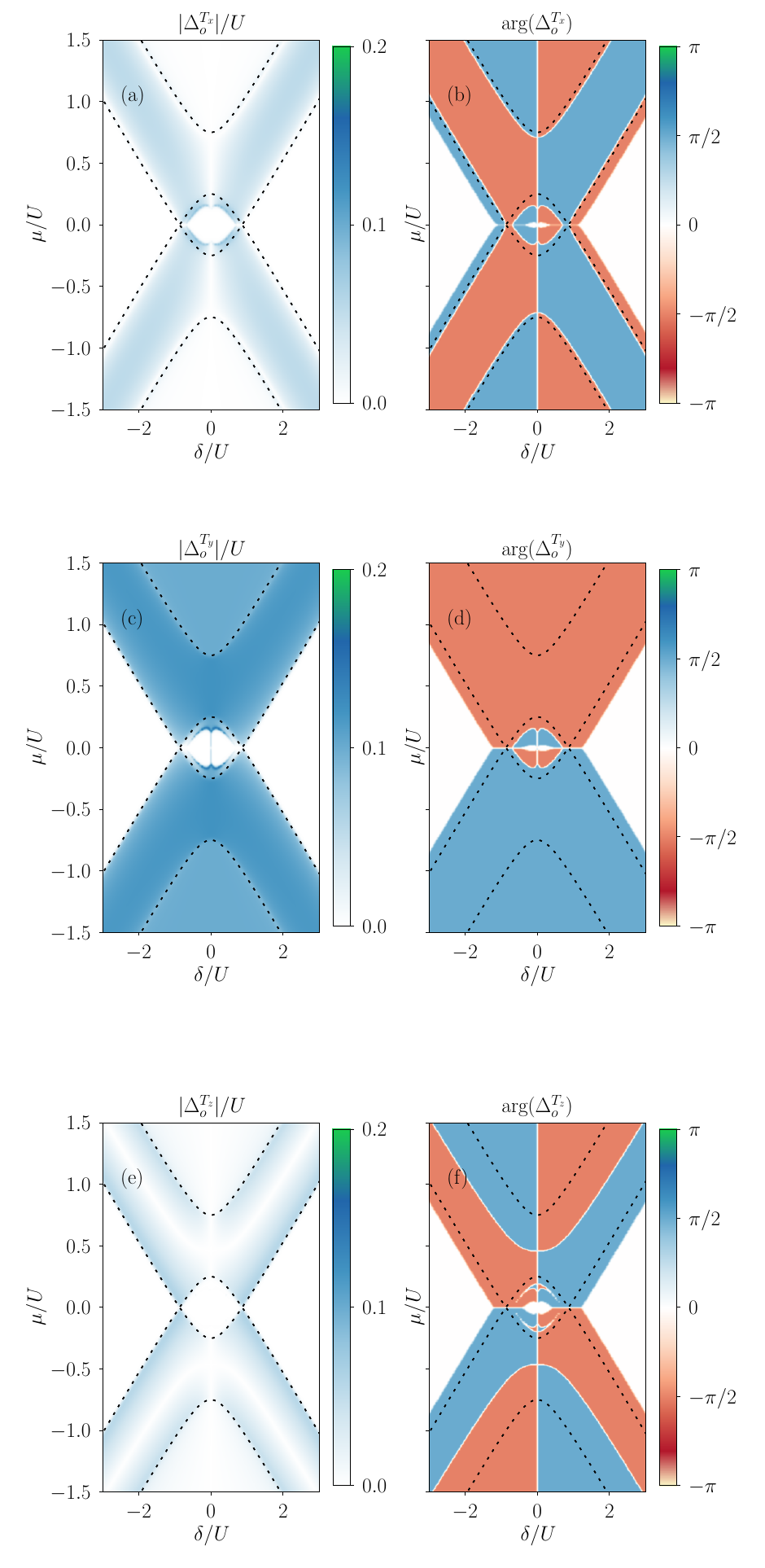}
\caption{\label{fig:fig16}(Color online) (a) Modulus and (b) phase of the odd-triplet order parameter for scenario (C) for parameters as in Fig.~\ref{fig:fig13}.}
\end{figure}

Finite Coulomb interaction at finite bias creates an exchange field that affects the accumulated spin on the dot. 
The resulting spin precession yields spin component not only within but also out of the plane defined by the leads magnetization directions.
Since the direction of the quantum dot spin is, in general, not collinear to the exchange field generated by the FM leads, the direction of the odd-triplet order parameter shows, according to Eq.~\eqref{delta_odd_relation}, a non-trivial dependence of the gate and bias voltage. 
This is illustrated in Fig.~\ref{fig:fig16} where we show the modulus and the phase of all components $\Delta_o^{T_x}, \Delta_o^{T_y}, \Delta_o^{T_z}$ of the odd-triplet order parameter.
Due to finite values of ${\bf S}_x, {\bf S}_y, {\bf S}_z$, all three components of $\mathbf \Delta_o$ are finite, see Eq.~\eqref{delta_odd_relation}.
For the chosen parameters, the phase of all order parameters is $\pm\pi/2$, indicating a purely imaginary odd triplet order parameter. 
Sign changes in the order parameter are due to a sign change of the spin, when going for instance from negative to positive gate voltage. We mention that all components are of the same order of magnitude due to a large spin-valve effect. In comparison, for $\phi_L=\phi_R$ all order parameters vanish, since the exchange field vanishes and spin symmetry is restored for the case $\mu_L=-\mu_R$. The positions of the resonances are determined by Andreev addition energies. 

To summarize scenario (C), spin symmetry can be completely removed by using two, non-collinearly magnetized FM leads.
The combination of spin-accumulation due to a voltage bias between the two ferromagnets and and spin-precession due to exchange fields yields finite values of all three components of the odd-triplet order parameter.
When tuning gate voltages, it is possible to flip the spin and therefore changing the sign of the imaginary part of the order parameter. 

\section{Conclusions}
\label{sec:conclusions}
In conclusion, we have studied the role of spin symmetry in the presence of superconducting correlations in a single-level quantum dot.
Conventional, even-singlet correlations are induced by a tunnel coupling to a superconducting lead with ordinary, BCS pairing symmetry.
In addition, superconducting odd-triplet correlations can be generated by breaking the $SU(2)$ spin symmetry with the help of a Zeeman field or ferromagnetic leads.
In a single-level quantum dot, no other pairing symmetries (even triplet or odd singlet) are possible.

We have determined the superconducting order parameters by numerically solving the kinetic equations for the dot degrees of freedom.
Thereby, we have taken all coherences between different charge states and between different spin states into account.
We have discussed three different scenarios in order to investigate the influence of reduced spin symmetry on the superconducting correlations: (A) full $SU(2)$ spin symmetry, (B) residual $U(1)$ spin symmetry, and (C) full removal of spin symmetry.
We have found that odd-triplet correlations are present in the dot whenever the dot is coupled to a superconductor and either a finite spin is accumulated or a magnetic field is applied.
All induced order parameters display a strong dependence on applied gate and bias voltages, magnetic fields, and the magnetization directions of attached ferromagnetic leads. 
This opens the possibility to selectively tune the even-singlet and odd-triplet order parameters.
In particular, we found regimes in which the unconventional, odd-triplet, correlations are large while the conventional, even-singlet, correlations are suppressed.

\section{Acknowledgements}
We acknowledge financial support of the DFG under project WE $5733/1$ and KO $1987/5$. We thank P. Stegmann and B. Sothmann for discussions.


\begin{thebibliography}{apsrev4-1}
\bibitem{bcs} J. Bardeen, L.N. Cooper, and J.R. Schrieffer, {\it Theory of Superconductivity}, Phys. Rev.  {\bf 108}, 1175  (1957).
\bibitem{Fertig} W.A. Fertig, D.C. Johnston, L.E. DeLong, R.W. McCallum, M.B. Maple, and B.T. Matthias, {\it Destruction of Superconductivity at the Onset of Long-Range Magnetic Order in the Compound ErRh$_4$B$_4$}, Phys. Rev. Lett. {\bf 38}, 987 (1977).
\bibitem{Ishikawa} M. Ishikawa and O. Fischer, {\it Magnetic ordering in the superconducting state of rare earth molybdenum sulphides, (RE)$_{1.2}$Mo${_6}$S${_8}$ (RE = Tb, DyandEr)}, Solid State Comm. {\bf 24}, 747 (1977).
\bibitem{Fulde} P. Fulde and R.A. Ferrell, {\it Superconductivity in a Strong Spin-Exchange Field}, Phys. Rev. {\bf 135}, A550 (1964).
\bibitem{Larkin} A.I. Larkin and Yu.N. Ovchinnikov, Zh. Eksp. Teor. Fiz. {\bf 47} 1136 (1964); Sov. Phys. JETP {\bf 20}, 762 (1965).
\bibitem{Eschrig_rep}M. Eschrig, {\it Spin-Polarized Supercurrents for Spintronics: A Review of Current Progress}, Rep. Prog. Phys. {\bf 78}, 104501 (2015).
\bibitem{Halterman} K. Halterman, P.H. Barsic, and O.T. Valls, {\it Odd Triplet Pairing in Clean Superconductor/Ferromagnet Heterostructures}, Phys. Rev. Lett. {\bf 99} 127002 (2007).
\bibitem{Fritsch} D. Fritsch and J.F. Annett, {\it Proximity effect in superconductor/conical magnet/ferromagnet heterostructures}, New J. Phys. {\bf 16}, 055005 (2014).
\bibitem{bergeret_rmp} F.S. Bergeret, A.F. Volkov, and K.B. Efetov, {\it Odd triplet superconductivity and related phenomena in superconductor-ferromagnet structures}, Rev. Mod. Phys. {\bf 77}, 1321 (2005).
\bibitem{leggett_theoretical_1975} A.J. Leggett,  {\it A theoretical description of the new phases of liquid 
$^3$He}, Rev. Mod. Phys. {\bf 47}, 331 (1975).
\bibitem{Ryazanov} V.V. Ryazanov, V.A. Oboznov,   A.Y. Rusanov,  A.V. Veretennikov, A.A. Golubov, and J. Aarts, {\it Coupling of Two Superconductors through a Ferromagnet: Evidence for a $\pi$-Junction}, Phys. Rev. Lett. {\bf 86}, 2427 (2001).
\bibitem{Kontos2002} T. Kontos, M. Aprili, J. Lesueur, G. Genet, B. Stephanidis, and B. Boursier, {\it Josephson Junction through a Thin Ferromagnetic Layer: Negative Coupling}, Phys. Rev. Lett. {\bf 89}, 137007 (2002).
\bibitem{robinson} J.W.A. Robinson, J.D.S. Witt, and M.G. Blamire, {\it Controlled Injection of Spin-Triplet Supercurrents into a Strong Ferromagnet}, Science {\bf 329}, 59 (2010).
\bibitem{birge} T.S. Khaire, M.A. Khasawneh, W.P. Pratt, Jr., and N.O. Birge, {\it Observation of Spin-Triplet Superconductivity in Co-Based Josephson Junctions}, Phys. Rev. Lett. {\bf 104}, 137002 (2010).
\bibitem{keizer} R.S. Keizer, S.T.B. Goennenwein, T.M. Klapwijk, G. Miao, G. Xiao, and A. Gupta, {\it A spin triplet supercurrent through the half-metallic ferromagnet CrO$_2$},Nature {\bf 439}, 825 (2006).
\bibitem{Bergeret_conv} F.S. Bergeret and I.V. Tokatly, {\it Singlet-Triplet Conversion and the Long-Range Proximity Effect in Superconductor-Ferromagnet Structures with Generic Spin Dependent Fields}, Phys. Rev. Lett. {\bf 110}, 117003 (2013).
\bibitem{tsuei_pairing_2000} C.C. Tsuei and J.R. Kirtley, {\it Pairing symmetry in cuprate superconductors}, Rev. Mod. Phys. {\bf 72}, 969 (2000).
\bibitem{mackenzie_superconductivity_2003} A.P. Mackenzie and Y. Maeno, {\it The superconductivity of Sr$_2$RuO$_4$ and the physics of spin-triplet pairing}, Rev. Mod. Phys. {\bf 75}, 657 (2003).
\bibitem{Martinez} W.M. Martinez, W.P. Pratt, Jr., and N.O. Birge, {\it Amplitude Control of the Spin-Triplet Supercurrent in $S/F/S$ Josephson Junctions}, Phys. Rev. Lett. {\bf 116}, 077001 (2016).
\bibitem{Gingrich_2015} E.C. Gingrich, B.M. Niedzielski, J.A. Glick, Y. Wang, D.L. Miller, R. Loloee, W.P. Pratt Jr., and N.O. Birge, {\it Controllable $0$-$\pi$ Josephson junctions containing a ferromagnetic spin valve}, Nat. Phys. {\bf 12}, 564 (2016).
\bibitem{Klose} C. Klose, T.S. Khaire, Y. Wang, W.P. Pratt, N.O. Birge, B.J. McMoran, T.P. Ginley, J.A. Borchers, B.J. Kirby, B.B. Maranville, and J. Unguris, Phys, {\it Optimization of Spin-Triplet Supercurrent in Ferromagnetic Josephson Junctions}, Rev. Lett {\bf 108}, 127002 (2012).
\bibitem{super_spintronics} J. Linder and J.W.A. Robinson, {\it Superconducting spintronics}, Nat. Phys. {\bf 11}, 307 (2015).
\bibitem{senapati}K. Senapati, M.G. Blamire, and Z.H. Barber, {\it Spin-filter Josephson junctions}, Nat. Mater. {\bf 10}, 849 (2011).
\bibitem{Tanaka} Y. Tanaka and A.A. Golubov, {\it Theory of the Proximity Effect in Junctions with Unconventional Superconductors}, Phys. Rev. Lett. {\bf 98}, 037003 (2007).
\bibitem{Balatsky} A. Balatsky and E. Abrahams, {\it New class of singlet superconductors which break the time reversal and parity}, Phys. Rev. B {\bf 45}, 13125 (1992).
\bibitem{de_franceschi_hybrid_2010} S. De Franceschi, L. Kouwenhoven, C. Sch\"onenberger, and W. Wernsdorfer, {\it Hybrid superconductor-quantum dot devices}, Nat. Nano. {\bf 5}, 703 (2010).
\bibitem{martin-rodero_josephson_2011} A. Mart\'in-Rodero and A. Levy Yeyati, {\it Josephson and Andreev transport through quantum dots}, Advances in Physics {\bf 60}, 899 (2011).
\bibitem{Schonenberger} L. Hofstetter, S. Csonka, J. Nygard, and C. Sch\"onenberger, {\it Cooper pair splitter realized in a two-quantum-dot Y-junction}, Nature {\bf 461}, 960 (2009).
\bibitem{Hermann} L.G. Herrmann, F. Portier, P. Roche, A. Levy Yeyati, T. Kontos, and C. Strunk, {\it Carbon Nanotubes as Cooper-Pair Beam Splitters}, Phys. Rev. Lett. {\bf 104}, 026801 (2010).
\bibitem{Heiblum} A. Das, Y. Ronen, M. Heiblum, D. Mahalu, A.V. Kretinin, and H. Shtrikman, {\it High-efficiency Cooper pair splitting demonstrated by two-particle conductance resonance and positive noise cross-correlation}, Nat. Comm. {\bf 3}, 1165 (2012).
\bibitem{Hussein2016} R. Hussein, L. Jaurigue, M. Governale, and A. Braggio, {\it Double quantum dot Cooper-pair splitter at finite couplings}, Phys. Rev. B {\bf 94}, 235134 (2016).
\bibitem{WeissUnconv} B. Sothmann, S. Weiss, M. Governale, and J. K\"onig, {\it Unconventional Superconductivity in Double Quantum Dots},  
Phys. Rev. B {\bf 90}, 220501(R) (2014).
\bibitem{BjoernMajo}O. Kashuba, B. Sothmann, P. Burset, and B. Trauzettel, {\it Majorana STM as a perfect detector of odd-frequency superconductivity}, Phys. Rev. B {\bf 95}, 174516 (2017).
\bibitem{Kontos} A.D. Crisan, S. Datta, J.J. Viennot, M. R. Delbecq, A. Cottet, and  T. Kontos, {\it Harnessing spin precession with dissipation}, Nat. Comm. {\bf 7}, 10451 (2015).
\bibitem{fazio98}
 R. Fazio and R. Raimondi, {\it Resonant Andreev Tunneling in Strongly Interacting Quantum Dots}, Phys. Rev. Lett. \textbf{80}, 2913 (1998).
\bibitem{schwab99}
P. Schwab and R. Raimondi, {\it Andreev tunneling in quantum dots: A slave-boson approach}, Phys. Rev. B \textbf{59}, 1637 (1999).
\bibitem{clerk00}
A.A. Clerk, V. Ambegaokar, and S. Hershfield, {\it Andreev scattering and the Kondo effect}, Phys. Rev. B \textbf{61}, 3555 (2000).
\bibitem{cuevas01}
J.C. Cuevas, A. Levy Yeyati, and A. Mart\'in-Rodero, {\it Kondo effect in normal-superconductor quantum dots}, Phys. Rev. B \textbf{63}, 094515 (2001).
\bibitem{avishai03}
Y. Avishai, A. Golub, and A.D. Zaikin, {\it Superconductor-quantum dot-superconductor junction in the Kondo regime}, Phys. Rev. B \textbf{67}, 041301 (2003).
\bibitem{dell-anna08}
L. Dell`Anna, A. Zazunov, and R. Egger, {\it Superconducting nonequilibrium transport through a weakly interacting quantum dot}, Phys. Rev. B \textbf{77}, 104525 (2008).
\bibitem{koerting10}
V. Koerting, B.M. Andersen, K. Flensberg, and J. Paaske, {\it Nonequilibrium transport via spin-induced subgap states in superconductor/quantum dot/normal metal cotunnel junctions}, Phys. Rev. B \textbf{82}, 245108 (2010).
\bibitem{pala07}
M.G. Pala, M. Governale, and J. K\"onig, {\it Non-Equilibrium Josephson and Andreev Current through Interacting Quantum Dots}, New. J. Phys. \textbf{9}, 278 (2007); \textbf{10}, 099801 (2008).
\bibitem{governale08}
M. Governale, M.G. Pala, and J. K\"onig, {\it Real-Time Diagrammatic Approach to Transport through Interacting Quantum Dots with Normal and Superconducting Leads }, Phys. Rev. B \textbf{77}, 134513 (2008); \textbf{78}, 069902 (2008). 
\bibitem{rozhkov00}
A.V. Rozhkov and D.P. Arovas, {\it Interacting-impurity Josephson junction: Variational wave functions and slave-boson mean-field theory}, Phys. Rev. B \textbf{62}, 6687 (2000).
\bibitem{tanaka07}
Y. Tanaka, A. Oguri, and A.C. Hewson, {\it Kondo effect in asymmetric Josephson couplings through a quantum dot}, New J. Phys. \textbf{9}, 115 (2007).
\bibitem{karrasch08}
C. Karrasch, A. Oguri, and V. Meden, {\it Josephson current through a single Anderson impurity coupled to BCS leads}, Phys. Rev. B \textbf{77}, 024517 (2008).
\bibitem{Sothmann} B. Sothmann, D. Futterer, M. Governale, and J. K\"onig, {\it Probing the exchange field of a quantum-dot spin valve by a superconducting lead}, Phys. Rev. B {\bf 82}, 094514 (2010).
\bibitem{Braggio} A. Braggio, M. Governale, M.G. Pala, and J. K\"onig, {\it Superconducting proximity effect in interacting quantum dots revealed by shot noise}, Solid State Comm. {\bf 151}, 155 (2011).
\bibitem{David} D. Futterer, J. Swiebodzinski, M. Governale, and J. K\"onig, {\it Renormalization effects in interacting quantum dots coupled to superconducting leads}, Phys. Rev. B {\bf 87}, 014509 (2013).
\bibitem{Eldridge} J. Eldridge, M. Pala, M. Governale, and J. K\"onig, {\it Superconducting proximity effect in interacting double-dot systems}, Phys. Rev. B {\bf 82}, 184507 (2010). 
\bibitem{QDSV_1} J. K\"onig and J. Martinek, {\it Interaction-Driven Spin Precession in Quantum-Dot Spin Valves}, Phys. Rev. Lett. {\bf 90}, 166602 (2003).
\bibitem{QDSV_2} M. Braun, J. K\"onig, and J. Martinek, {\it Theory of transport through quantum-dot spin valves in the weak-coupling regime}, Phys. Rev. B {\bf 70}, 195345 (2004).
\bibitem{foot1} Since $F_{\sigma\sigma'}(-t)=\langle T d_\sigma(-t)d_{\sigma'}(0)\rangle=\langle Td_\sigma(0)d_{\sigma'}(t)\rangle=-\langle T d_{\sigma'}(t)d_\sigma(0)\rangle=-F_{\sigma'\sigma}(t)$, it follow $F_e^S(t)=F_e^S(-t)$ and correspondingly for $F_o^{T_i}(-t)=-F_o^{T_i}(t)$ with $i=x,y,z$.  
\bibitem{balatsky_even-_1993} A.V. Balatsky and J. Bon\u{c}a, {\it Even- and odd-frequency pairing correlations in the one-dimensional t-J-h model: A comparative study}, Phys. Rev. B {\bf 48}, 7445 (1993).
\bibitem{Braun_noise} M. Braun, J.K\"onig, and J. Martinek, {\it  Frequency-dependent current noise through quantum-dot spin valves}, Phys. Rev. B {\bf 74}, 075328 (2006).
\bibitem{Buettiker_cross} M. B\"uttiker, {\it Role of scattering amplitudes in frequency-dependent current fluctuations in small conductors}, Phys. Rev. B {\bf 45}, 3807 (1992).
\bibitem{Martin_1992} Th. Martin and R. Landauer, {\it Wave-packet approach to noise in multichannel mesoscopic systems}, Phys. Rev. B {\bf 45}, 1742 (1992).

\end{thebibliography}
\end{document}